\documentclass[12pt]{article}

\usepackage[english]{babel}
\usepackage[T1]{fontenc}
\usepackage[latin2]{inputenc}
\usepackage[centertags]{amsmath}
\usepackage{amssymb}
\usepackage{amsthm}
\usepackage{graphicx}
\input xy
\xyoption{all}

\voffset=-2cm\date{} 

\newtheorem{twr}{Theorem}
\newtheorem{prop}{Proposition}
\newtheorem{lem}{Lemma}
\newtheorem{wn}{Corollary}
\newtheorem{rem}{Remark}
\title{ \bf Growth of polyhedral crystals \\from supersaturated vapor}
\author{ \bf Przemysław Górka \\ Department of Mathematics and Information Sciences,\\ Warsaw University of Technology,\\
Pl. Politechniki 1, 00-661 Warsaw, Poland \\ e-mail: \texttt{pgorka@mini.pw.edu.pl}
           }

\begin{document}
\date{}
\maketitle
\begin{abstract} We examine the growth of crystals from vapor. We assume
that the~Wulff shape is a~prism with a~hexagonal base.
The~Gibbs-Thomson correction on the~crystal surface is included in
the model. Assuming that the-initial crystal has an admissible shape
 we show local in time existence
 of solutions.
\end{abstract}


\section{Introduction}
The main result of this paper is a~mathematical study of crystals
grown from supersaturated vapor. We assume that evolving crystal
$\Omega$ is a~prism with $N$-gonal base. We assume that the~number $N$
is constant. Let us stress that our crystal $\Omega$ has not to be
a convex set (see Figure~1 in the case $N=54$). Such kind of ice crystals
are formed in the atmosphere.
Our activity is motivated by Gonda and Gomi results,
\cite{Gonda}, and~also by Giga and~Rybka \cite{Giga1}. The authors
of the~paper \cite{Giga1} assumed that the~Wulff shape is a~fixed
cylinder and~also that the~process was slow, i.e. they considered
quasi-steady approximation of~ the diffusion equation (diffusion
is much faster than the evolution of free boundary). Giga and
Rybka obtained results for quasi-steady approximation system (see
\cite{Giga2}, \cite{Giga3}, \cite{Giga4}, \cite{Giga5}).
\begin{center}
    \includegraphics[scale=.25]{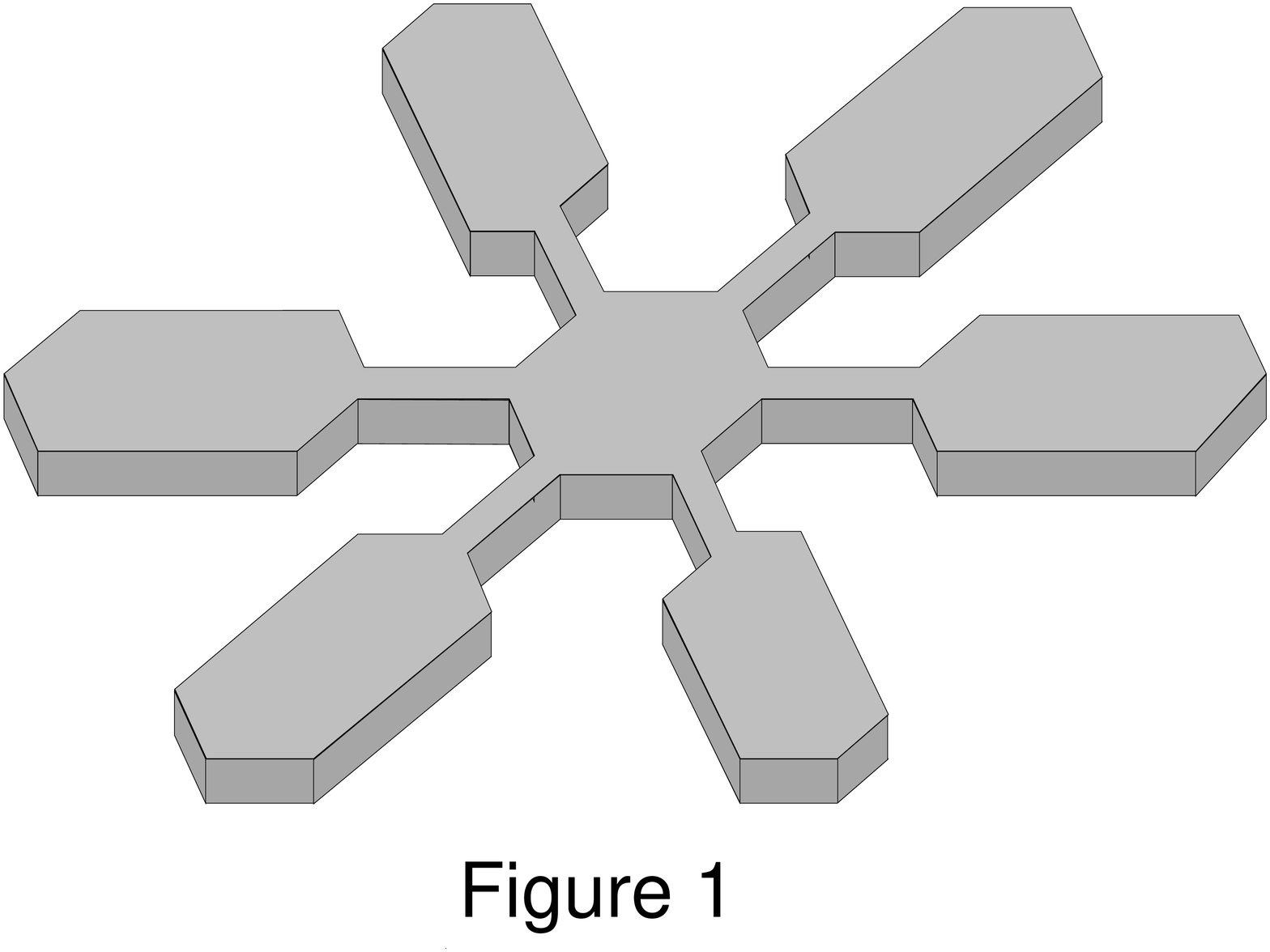}
\end{center}
In this paper we do not assume that~the~crystal evolves slowly.
Namely, we assume that~supersaturation  $\sigma$ fulfills
diffusion equation with a~drift
\begin{equation}
    \label{dyfuzja}
    \varepsilon \frac{\partial}{\partial t} \sigma=\Delta \sigma + f\cdot \nabla \sigma
\end{equation}
outside of crystal $\Omega$. For simplicity of notation we assume
that $\varepsilon=1$. We are not going to~examine the properties
of the~solution when we change the~parameter $\varepsilon$. In
the~above equation $f$~is~a~velocity of~vapor. We assume that
$f$~is~given. In fact, velocity $f$ should satisfy an addition
equation.

We assume that the supersaturation $\sigma$ has a~specific value at infinity, i.e.
\begin{eqnarray}
    \label{nieskonczonosc}
    \lim_{|x| \rightarrow \infty} \sigma (x) = \sigma^{\infty}.
\end{eqnarray}
We require that mass conservation law on the~crystal surface is fulfilled. Namely,
\begin{eqnarray}
      \label{brzeg}
       g=\frac{\partial\sigma}{\partial \nu},
\end{eqnarray}
where $\nu$ is the~outer normal. This condition is very natural in
the epitaxy model. See Eliot and  Gurtin paper \cite{Eliot}. Let
us stress that it could be better to discussed the system with
replaced $g$ by $V$ in (\ref{brzeg}), where  $V$ is the~velocity
of the~growing crystal. Namely, $V=\frac{\partial\sigma}{\partial
\nu}$. This is the~so-called Stefan condition. Our method can not
be applied to this case. This problem will be examined in the
forthcoming article \cite{Gorka}.

The value of the supersaturation $\sigma$ at the~surface satisfies
Gibbs-Thomson law, namely,
\begin{eqnarray}
          \label{g-t}
          -\sigma=-div_{S}\xi-\beta V,
\end{eqnarray}
where $\beta$ is the~kinetic coefficient and $\xi$ is
a~Cahn-Hoffman vector field (see \cite{Gurtin}). We denote by $V$
the~velocity of the~growing crystal. We also assume
that~the~velocity $V$ is constant on each side of the~crystal. Our
model does not include bending and breaking of surface $\partial
\Omega$. Roughly speaking the~above relation says
the~supersaturation on the~crystal surface is proportional to~the
curvature of surface ($-div_{S}\xi$), and to the velocity of
the~evolving crystal. Let us discuss the~properties of
the~Cahn-Hoffman vector field $\xi$. Let $\gamma$ be surface
energy density (Finsler metric, see \cite{Rund}), i.e. $\gamma
:\mathbb{R}^3 \rightarrow \mathbb{R}$ is a~$1$-homogeneous, convex
and Lipschitz continuous function. If the surface of the~crystal
$\partial \Omega$ and the surface energy density $\gamma$ are
smooth, then
\begin{eqnarray}
\label{polech}
              \xi(x) = \nabla \gamma(\nu(x)).
\end{eqnarray}
If we do not assume that $\gamma$ and $\partial \Omega$ are smooth
then the above expression  may not make sense
(see~\cite{Paolini}). Instead $\nabla \gamma$ we can write
subdifferential $\partial \gamma$, which is defined everywhere. In
this case  $\partial \gamma$ is a~nonempty convex set (not
necessarily singleton). The condition (\ref{polech}) can be
replaced by the~following relation
\begin{eqnarray*}
              \xi(x) \in  \partial \gamma(\nu(x)).
\end{eqnarray*}
We are not going to discuss the properties of the~Cahn-Hoffman
vector field. In preliminaries section we~show a~"nice" property
of the~averaged divergence of the~Cahn-Hoffman field $\xi$, namely
\begin{displaymath}
        -\kappa_i\mathcal{H}^2(S_i)=\int_{S_i}div_{S}\xi \, d  \mathcal{H}^2(x),
\end{displaymath}
where $\kappa_i$ is a crystalline curvature (see next section).
Let us stress that the LHS of the above expression is independent
of  $\xi$.

In this paper we assume that the~Frank diagram  $F_{\gamma}$ (see
\cite{Gurtin} or next section) is a~sum of two regular pyramids
having a~common base. Hence, the~Wulff shape $W_{\gamma}$ is
a~prism with a~hexagonal base. We assume that the~crystal is an
admissible shape, i.e. the~set of outer normal vectors to
$\partial \Omega$ coincides with the~set of outer normal vectors
to $W_{\gamma}$. Let us notice that $\Omega$ need not be a~convex
set. Let us stress that we could apply our analysis to more general Wulff shape. Namely,
we could assume that $W_{\gamma}$ is a~prism with regular $k$-gonal base, where $k \geq 3$.
But only hexagonal symmetry is intersting from pysical point of view (see Figure 1).

In order to simplify our system (\ref{dyfuzja}),
(\ref{nieskonczonosc}), (\ref{brzeg}), (\ref{g-t}) we will
consider averaged Gibbs-Thomson law. Taking into account the above
considerations and the assumption that
 the~velocity~$V$ is~constant on each side of~the~crystal we~obtain
\begin{eqnarray}
    \label{sgt}
    -\int_{S_i} \sigma  d  \mathcal{H}^2 =\kappa_i\mathcal{H}^2(S_i) - \beta_i V_i \mathcal{H}^2(S_i).
\end{eqnarray}
Let us mention that averaged Gibbs-Thomson is not new condition.
Namely, this equation has appeared in \cite{Gurtin1} and
\cite{Herring}.

Our main aim is to show the local in time existence of solutions
to (\ref{dyfuzja}), (\ref{nieskonczonosc}), (\ref{brzeg}),
(\ref{sgt}). We show this in~a~few steps. Let us notice that we
can look at this system as parabolic equation coupled with
ordinary differential equation for the evolving sides of~crystal
(the signed distance side from $S_i(t)$ to $S_i(0)$). Next, we
transform our problem to the~system in the fixed set, in this case
it is outside the initial crystal. In this domain we solve our
system. Subsequently, we examine the linear problem and here we
apply the analytic semigroup theory (see \cite{Ac}, \cite{Ac1},
\cite{Amann}, \cite{Engel}, \cite{Henry}, \cite{Lunardi}). Next we
construct an approximating sequence and apply Ascoli's Theorem.
Let us notice that we work in an unbounded domain and the standard
Rellich-Kondrachov compactness theorem does not work. We have to
show additional properties of the approximating sequence. Our
method is similar to the method presented in \cite{Rybka}.

The paper is organized as follows. First, we introduce notations,
recall definitions, show some facts about crystalline curvature
and formulate the problem. Next, we show the~main result. The~last
section (Appendix) contains useful result.

At the~end of this section, let us comment on the mathematical
literature. In the~smooth case, i.e. assuming that the surface of
the~crystal is differentiable manifold and for the two phase
Stefan problem this model was solved in~the~class of smooth
function (see Chen and Reitich \cite{Chen}). Independently,
Radkevich (see \cite{Radkevich}) showed local in time existence of
a smooth solution to the problem considered by Chen and Reitich.
The advantage of Radkevich is that the author allows a slightly
more general form of the diffusion equation. The problem for
$\beta =0$ and smooth interfaces was studied by Luckhaus
\cite{Luckhaus}
 and in greater generality by Almgren-Wang \cite{Almgren}. In
 particular, they showed that uniqueness fails. Let us stress that
 these authors have worked in bounded domain.
 Quasi-steady approximation in~the~case when the crystal is
 a~cylinder and a domain is unbounded
was discussed by Giga and Rybka \cite{Giga1}. Let us mention that
these authors discussed the system with replaced $g$ by $V$ in
(\ref{brzeg}), i.e. the~so-called Stefan condition. Our
method can not be applied to this case.\\
\\
\textbf{Notation.}  We use the
following convention, whenever we see the inequality $\mathbb{A}
\leq c \mathbb{B}$ we tacitly understand that it holds with some
positive constant $c$ independent of~$\mathbb{A}$ and
$\mathbb{B}$.
\section{Preliminaries}

Let us denote an evolving crystal by $\Omega(t)$, and its exterior
by $\Omega(t)^c =\mathbb{R}^3 \setminus \Omega(t)$. Let $S(t)
=\partial \Omega(t)$ its surface. We assume that crystal
$\Omega(t)$ is a~prism at all times. To be more precise we
write
\begin{eqnarray*}
    \Omega (t)= \{(\tilde{x_1},\tilde{x_2},\tilde{x_3})\in \mathbb{R}^3: L_{B}(t)\leq \tilde{x_3}
      \leq L_{T}(t), (\tilde{x_1},\tilde{x_2})\in
    W(t)\}.
\end{eqnarray*}
Where $W(t)$ is an $N$-gon in~the~plane.

Let us divide the surface into $N+2$ pieces, i.e.
\begin{eqnarray*}
    S_i(t)=\left\{(\tilde{x_1},\tilde{x_2},\tilde{x_3}):(\tilde{x_1},\tilde{x_2})\in L_i(t),L_{B}(t)\leq \tilde{x_3}  \leq L_{T}(t)\right\},\\
    S_T(t)=\{(\tilde{x_1},\tilde{x_2},\tilde{x_3}):(\tilde{x_1},\tilde{x_2})\in W(t),\tilde{x_3}= L_{T}(t)\},\\
    S_B(t)=\{(\tilde{x_1},\tilde{x_2},\tilde{x_3}):(\tilde{x_1},\tilde{x_2})\in W(t),\tilde{x_3}= L_{B}(t)\},
\end{eqnarray*}
where $i=1,2,...,N$ (we also use the notation $N+1 =T$ and
$N+2=B$). The~sets $S_i$ shall be called facets. In order to
understand this convention it is better to look at Figure 2.
\begin{center}
    \includegraphics[scale=.2]{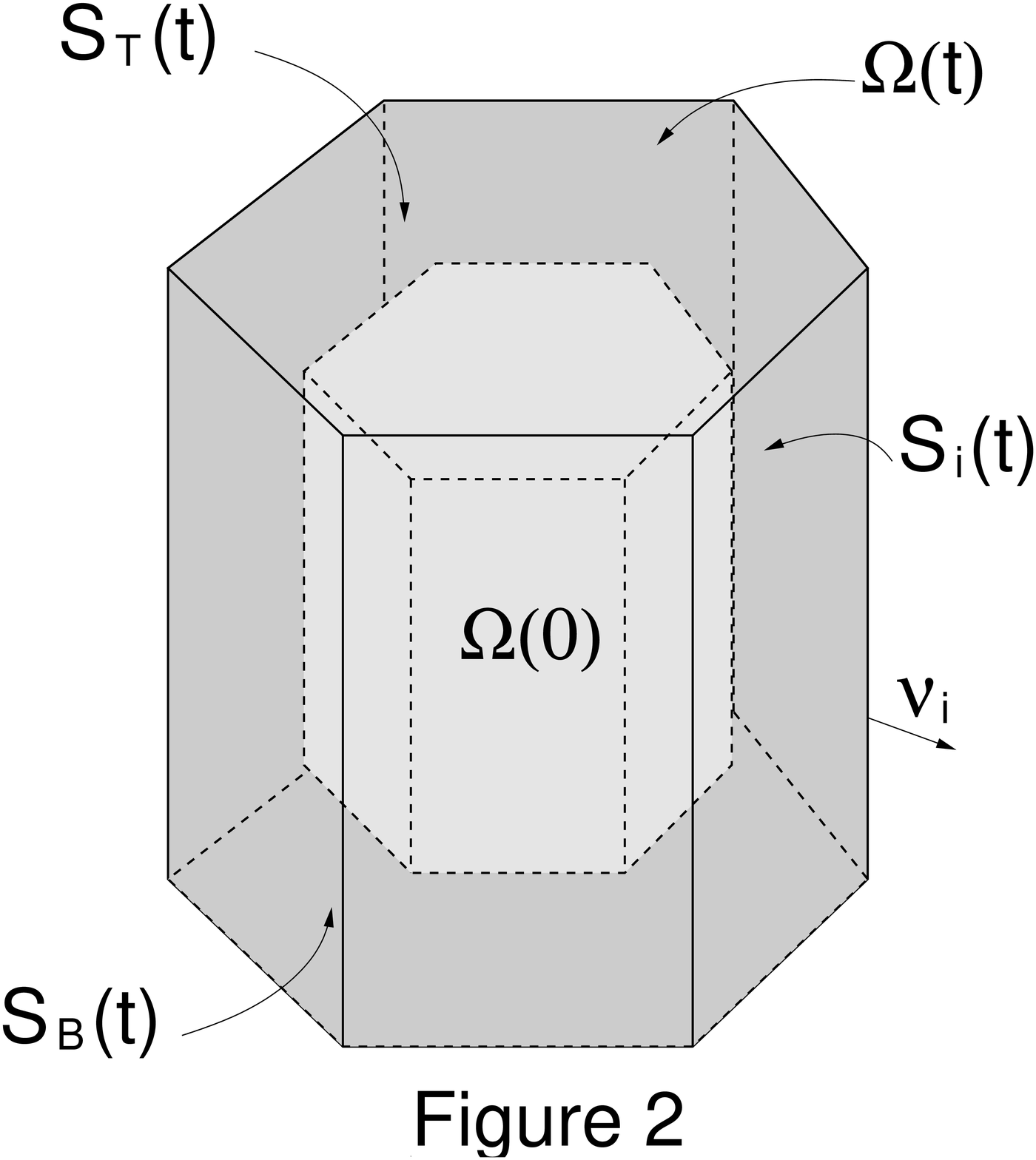}
\end{center}
Let us denote by $L(t)$ the~boundary in the plane of $N$-gon
$W(t)$, namely
\begin{eqnarray*}
   L(t)=\partial W(t)=\bigcup_{i=1}^{N}L_i(t),
\end{eqnarray*}
where $L_i(t)$ are edges. Let us additionally write $\Omega_0
=\Omega(0)$, and the~coordinate systems will be presented as
follows
\begin{eqnarray*}
    (\tau,x_1,x_2,x_3) \in \mathbb{R}_{+} \times \Omega^c_0\\
     (t, \tilde{x_1}, \tilde{x_2}, \tilde{x_3} )\in \{t\} \times \Omega^c(t).
\end{eqnarray*}
We denote by $n$ the~outer normal vector to the~boundary $\Omega_0^c$ and by $\nu$ the~outer normal to~$\Omega^c(t)$.

We denote the coordinates of vertices of $\Omega_0$ by
\begin{eqnarray*}
    \left(x_1^i,x_2^i,\frac{L}{2}\right) \text{ and } \left(x_1^i,x_2^i, -\frac{L}{2}\right),
\end{eqnarray*}
where $i=1,...,N$ and $L$ is the~height of $\Omega_0$. Through the
paper we shall assume that $0 \in \Omega(0)$. Analogously, we denote
the~coordinates of the vertices of $\Omega(t)$, i.e.
\begin{eqnarray*}
    \left(x_1^i(t),x_2^i(t),L_{T}(t)\right), \left(x_1^i,x_2^i,L_{B}(t)\right), \text{ where } i=1,...,N.
\end{eqnarray*}
Next, we denote by $z_i(t)$ the~signed~distance between
$i$-th~face of~$\Omega(0)$ and $i$-th face of $\Omega(t)$, i.e.
\begin{eqnarray*}
    z_i(t) &=&  \begin{cases}
            dist (S_i(t),S_i(0)), \quad \text{if}  \, \left(x_1^i(t) - x_1^i(0),x_2^i(t)-x_2^i(0),L_{T}(t)-L_{T}(0) \right) \cdot \nu_i
        >0, \\
                -dist(S_i(t),S_i(0)), \quad \text{if} \, \left(x_1^i(t) - x_1^i(0),x_2^i(t)-x_2^i(0),L_{T}(t)-L_{T}(0) \right) \cdot \nu_i <0.
            \end{cases}
\end{eqnarray*}
Making not difficult but boring and long calculations one can
examine what is the~dependence  between $z_i(t)$, $x_j(t)$ and
$x_k$. We will summarize them. We use the following convention
$x_j=x_j(t)$. Namely, let us denote by
\begin{eqnarray*}
    c_i=x_2^i-x_2^{i+1}    \text{, }\:  e_i=x_1^i (0)-x_1^i, \\
    d_i=x_1^i-x_1^{i+1}    \text{, } \: f_i=x_2^i (0)-x_2^i.
\end{eqnarray*}
Next, we denote by
\begin{eqnarray*}
    k_i=\left( \begin{array}{cc} d_i \\ c_i \end{array} \right),
    w_i=\left( \begin{array}{cc} e_i \\ f_i \end{array} \right)
\end{eqnarray*}
We can assume that $ x_1^i \neq x_1^{i+1}$ for $i=1,...,N$. Hence,
the~normal vector has the~form
\begin{eqnarray*}
    n_i=\frac{|x_1^i-x_2^i|}{\sqrt{(c_i)^2+(d_i)^2}}\left[\frac{c_i}{d_i},1\right]=[a_i,b_i].
\end{eqnarray*}
Finally, we can state the following lemma.
\begin{lem}
\begin{eqnarray*}
    x_1^i(t)=\frac{x_{1a_i}(t)}{m_i(t)},\: \:
    x_2^i(t)=\frac{x_{2a_i}(t)}{m_i(t)},
\end{eqnarray*}
where
\begin{eqnarray*}
x_{1ai}(t)=(x_1^i+a_{i-1}z_{i-1})(c_{i-1}+b_{i-1}z_{i-1})(d_i+a_iz_i)-(x_1^{i+1}+a_{i}z_{i})(c_{i}+b_{i}z_{i})(d_{i-1}+a_{i-1}z_{i-1})+\\
    -(x_2^i+b_{i-1}z_{i-1})(d_{i}+a_{i}z_{i})(d_{i-1}+a_{i-1}z_{i-1})+(x_2^{i+1}+b_{i}z_{i})(d_{i}+a_{i}z_{i})(d_{i-1}+a_{i-1}z_{i-1}),\\
    x_{2ai}(t)=(x_1^i+a_{i-1}z_{i-1})(c_{i}+b_{i}z_{i})(c_{i-1}+b_{i-1}z_{i-1})-(x_1^{i+1}+a_{i}z_{i})(c_{i}+b_{i}z_{i})(c_{i-1}+b_{i-1}z_{i-1})+\\
    +(x_2^{i+1}+b_{i}z_{i})(d_{i}+a_{i}z_{i})(c_{i-1}+b_{i-1}z_{i-1})-(x_2^i+b_{i-1}z_{i-1})(d_{i-1}+a_{i-1}z_{i-1})(c_{i}+b_{i}z_{i}),\\
    m_i(t)=(c_{i-1}+b_{i-1}z_{i-1})(d_{i}+a_{i}z_{i})-(c_{i}+b_{i}z_{i})(d_{i-1}+a_{i-1}z_{i-1}).
\end{eqnarray*}

\end{lem}
\begin{proof}
We leave those straightforward calculations to the interested reader.\\
\end{proof}

In our problem evolution is determined by normal vectors to~the
boundary of the initial crystal. It~means that~facets can move
parallel to the~facets of $\Omega_0$. We assume that the set of
normal vectors of $\Omega_0$ coincides with the set of normal
vectors of a~prism with a~hexagonal base. Such~$\Omega_0$ we~shall
call an admissible shape. Let us notice that we need $N+2$
parameters, in order to~describe the evolution of crystals, i.e.
distance between facets $z_i$. In our problem there appears
surface divergence $div_{S}\xi$. We define this quantity as
follows
\begin{eqnarray*}
    div_{S}\xi(x):=tr((Id-n(x) \otimes n(x))\nabla \xi(x)),
\end{eqnarray*}
where $n(x)$ is the unit vector orthogonal to~tangent space $T_x S$ (see \cite{Simon}).

\subsection{Crystalline curvature and surface energy}

Let us recall the basic objects from crystalline geometry, i.e.
Wulff shape $W_{\gamma}$ and Frank diagram $F_{\gamma}$. Let
$\gamma$ be a~surface energy density, i.e. $\gamma : \mathbb{R}^3
\rightarrow \mathbb{R}$ is a~$1$-homogeneous, convex and Lipschitz
continuous function. Then the Wulff shape and Frank diagram (see
\cite{Gurtin}) are defined as follows:
\begin{eqnarray*}
    W_{\gamma} = \left\{x \in \mathbb{R}^3 : \forall n \in \mathbb{R}^3, |n|=1, x \cdot n \leq \gamma (n)\right\}\\
    F_{\gamma} = \{x \in \mathbb{R}^3 : \gamma (x) \leq 1 \}.
\end{eqnarray*}
In our paper we assume that the Wulff crystal $W_{\gamma}$ is
a~prism with a~hexagonal base. This~assumption is~consistent with
physical experiments (see \cite{Petrenko} and Figure 1). Then, it is not hard
to~see that the Frank diagram $F_{\gamma}$ is a~sum of two regular
pyramids having a~common base.

Let us formulate the following
\begin{lem}
Let us assume $T_6 \in SO(2)$, i.e.
\begin{eqnarray*}
    T_6=\left( \begin{array}{ll} cos \frac{2\pi}{6} & -sin \frac{2 \pi}{6} \\
    sin \frac{2 \pi}{6} & cos \frac{2 \pi}{6} \end{array} \right)
\end{eqnarray*}
and $\gamma(1,0,0)=\gamma_L$. Let us also assume $T_6 (1,0)^{T} = (x_1',x_2')^{T}$ and
\begin{eqnarray*}
    \gamma_1 (\tilde x_1, \tilde x_2) = \gamma_L \left( \tilde x_1 - \left(\frac{x_1'}{x_2'}-\frac{1}{x_2'}\right) \tilde x_2 \right)
\end{eqnarray*}
and
\begin{eqnarray*}
    \gamma ( x_1,x_2,x_3) = \gamma_1 \left(T_6^{-(i-1)} (x_1,x_2)\right)+|x_3| \gamma_T \text{   if   } (x_1,x_2) \in A_i,
\end{eqnarray*}
where $A_i$ is $i$-th sector ($i=1,2,...6$), see Figure 3.
Then $F_{\gamma}$ is a~sum of two regular pyramid.
\begin{center}
    \includegraphics[scale=.4]{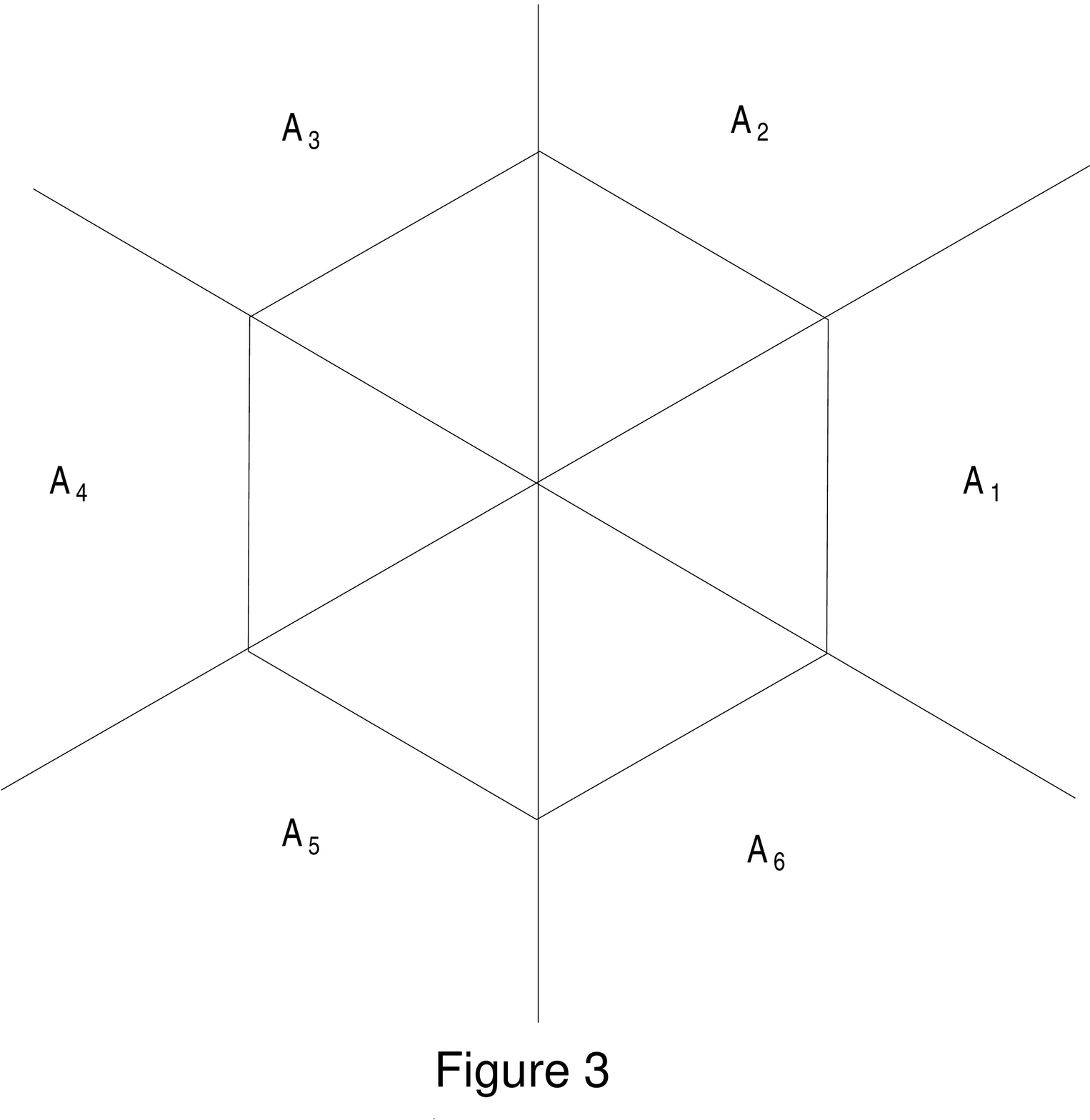}
\end{center}
\end{lem}
\begin{proof} In this relatively easy calculation homogenity of function $\gamma$ and elementary geometry have been
used. The details we left to the reader.
\end{proof}
Let us recall that the surface energy $E(S)$ is expressed by the formula:
\begin{eqnarray*}
    E(S)= \int_{S} \gamma (n(x)) d  \mathcal{H}^2(x),
\end{eqnarray*}
where $\mathcal{H}^2$ is two dimensional Hausdorff measure.

The crystalline curvature $\kappa_i$ is defined as (see \cite{Taylor}):
\begin{eqnarray*}
    \kappa_i=-\lim_{a \rightarrow 0} \frac{\Delta E}{\Delta V},
\end{eqnarray*}
where $a$ is the amount of motion of $S_i$ in the direction of the
outer normal to~$S$, $\Delta E$~is~the~resulting change of surface
energy, and $\Delta V$ is the change of volume. Let us denote by
$\alpha_i$ the~angle between $i$-th and  $i+1$-th facet. Now, we
show the~following lemma.
\begin{lem}
The crystalline curvature is given by expressions:
\begin{gather*}
    \kappa_T=\kappa_B= -\frac{\sum_{j=1}^{N}l_j \gamma(n_j)}{\mathcal{H}^2(S_T)}\\
    \kappa_j =-\frac{1}{l_j}\Big(ctg(\alpha_j)+ctg(\alpha_{j+1})\Big)\gamma(n_j) -\frac{1}{l_j}\frac{1}{\sin(\alpha_j)}\gamma(n_{j-1})
    -\frac{1}{l_j}\frac{1}{sin(\alpha_{j+1})}\gamma(n_{j+1})+\frac{2}{l}\gamma(n_T)
\end{gather*}
Where $l_j =|L_j|$ and $l=|L|$, $j=1,...,N$.
\end{lem}
\begin{proof}
It is a~straightforward calculation. It is not hard to~see that:
\begin{gather*}
 \Delta E_T  = \sum_{j=1}^{N} l_j a \gamma(n_j)\\
  \Delta V_T = a \mathcal{H}^2(S_T)\\
  \Delta E_j  =  la\Big(ctg(\alpha_j)+ctg(\alpha_{j+1})\Big)\gamma(n_j) + \\
   +  \frac{la}{sin(\alpha_j)} \gamma(n_{j-1})+\frac{la}{sin(\alpha_j+1)} \gamma(n_{j+1})+a\Big(2
   l_j+a\big(ctg(\alpha_j)+ctg(\alpha_{j+1})\big)\Big)\gamma(n_T)\\
   \Delta V_j  =
   \frac{a}{2}\Big(2l_j+a\big(ctg(\alpha_j)+ctg(\alpha_{j+1})\big)\Big)l,
\end{gather*}
where $j=1,...,N$. From those formulas the~claim follows.
\end{proof}

We finish this section with a theorem which let us average Gibbs-Thompson law.
 \begin{twr}\label{usrednianie}
The following equality is true:
\begin{displaymath}
    \kappa_i=-\frac{1}{\mathcal{H}^2(S_i)}\int_{S_i}div_{S}\xi d  \mathcal{H}^2(x).
\end{displaymath}
\end{twr}
\begin{proof}
    It is a similar calculation to~the proof of Proposition 2.1 from \cite{Giga4},
    where the Gauss formula was the main tool.
     We~leave this to~the~reader.
\end{proof}
Finally, we will work with the~system
\begin{eqnarray}
    \label{uklad}
    \frac{\partial}{\partial t} \sigma=\Delta \sigma + f\cdot \nabla \sigma  \text{  in  } \Omega^c(t)\nonumber\\
    g=\frac{\partial\sigma}{\partial \nu} \text{  on  } \partial \Omega^c(t)\\
            \frac{1}{\mathcal{H}^2(S_i(t))} \int_{S_i(t)} \sigma (x) \mathcal{H}^2(x) =-\kappa_i(t)+\beta_i V_i \nonumber,
\end{eqnarray}
where $g=(g_1,...,g_{N+2})$ is given.

\section{The existence of a solution - preparation}
Let us mention that we can assume that $\sigma^{\infty} =0$. At
the~beginning we transform the system to~the one with homogeneous
boundary conditions. From Lemma \ref{rozszerzenie} (see Appendix)
we know that there exists a~family of functions~$h_i$, where
$i=1,2...,N+2$ with compact supports, such that
\begin{eqnarray*}
    \frac{ \partial h_i} {\partial \nu_j } \Big|_ { { S_j (0)}} = \delta_{i,j}.
\end{eqnarray*}
Subsequently, let us introduce the notation
\begin{eqnarray*}
    G= \sum_{i=1}^{N+2} h_i g_i,
\end{eqnarray*}
and $u=\sigma-G$. Now we obtain the~system for $u$ with
homogeneous boundary conditions. Namely, we will deal with
the~following final problem
\begin{eqnarray}
    \label{problemu}
    \frac{\partial u}{\partial t} =\Delta u + f\cdot \nabla u +\left(\Delta + f \cdot \nabla -\frac{\partial}{\partial t}\right)G \text{  in  } \Omega^c(t),\nonumber\\
    \frac{\partial u}{\partial \nu} =0 \text{  on  } \partial \Omega^c(t),\\
     \frac{1}{\mathcal{H}^2(S_i(t))} \int_{S_i(t)} \big(u(x)+G \big)\,d \mathcal{H}^2(x) =-\kappa_i(t)+\beta_i
     V_i,    \nonumber\\
    u(0)=u_0, \, z(0)=0 \nonumber.
\end{eqnarray}
We assume that $f \in L^{\infty} (\mathbb{R}^3 \times
\mathbb{R}_{+} )$.
\subsection{Transformation of the~system to~the fixed domain}

In this subsection we transform our problem into a system in
a~fixed domain. In this case it is the outside of the initial
crystal $\Omega_0$. We have to~construct a family of
diffeomorphisms from $\Omega^c$ to~$ \Omega^c(t)$, i.e.
\begin{equation*}
    \Lambda : \Omega^c \rightarrow \Omega^c(t).
\end{equation*}
Let us notice, that $\Omega^c(t)$ depends on $z(t)$, namely
$\Omega^c(t)=\Omega^c(z(t))=\Omega^c(z)$.  We have the~following
main theorem of this section.
\begin{twr}
\label{dif} There exists a~family of diffeomorphisms $\Lambda_z$ of
$C^{2+ \alpha}$ class ($\alpha >0$)  transforming $\Omega^c(0)$ on
$\Omega^c(z)$. Additionally, the~family has the~following property,
if $|x_1,x_2,x_3| > 10 \hbox{ diam} (\Omega^c(z))$ then $\Lambda_z =
Id$. In addition, the dependence upon $z$ is smooth.
\end{twr}
Before we go to~the proof we make some comments.
\begin{rem}
If we make a~cross-section parallel to~$S_T$, then one of
the~problems is to find a~diffeomorphism transforming the exterior
of $N$-gon onto exterior another $N$-gon. It seems that we could
use complex analysis methods and apply Riemann Theorem. But here
we meet the~following problem, namely we work in
nonsimplyconnected domain. There exist the so-called
Christoffel-Schwarz expressions transforming half plane onto
interior of $N$-gon. The~second problem is the requirement of
smoothness of $\Lambda$ with respect to~$t$.
\end{rem}
\begin{rem}
The number $10$ in formulation of the theorem is not substantial.
We could take any~number greater than~$1$.
\end{rem}
\begin{rem}
During the examination of the transformed system we use the property
that $\Lambda_z$ is identity far from $S(z)$.
\end{rem}
\begin{proof}
We divide the construction into two steps, namely
\begin{eqnarray*}
    \Phi : \Omega^c \rightarrow \Sigma (z)\\
    \Psi : \Sigma(z) \rightarrow \Omega^c(z),
\end{eqnarray*}
where $\Sigma(z)$ is defined as follows:
\begin{equation*}
    \Sigma(z)=  \big\{ (x_1,x_2,x_3)\in \mathbb{R}^3: L_{B}(z)\leq x_3  \leq L_{T}(z), (x_1,x_2)\in W(0) \big\} ^c.
\end{equation*}
We define the map $\Lambda$ in the~following way
\begin{equation*}
    \Lambda= \Psi \circ \Phi.
\end{equation*}
First we define transformation $\Phi$ as follows
\begin{displaymath}
     \Phi (x_1,x_2, x_3)=\left\{ \begin{array}{ll} \Big(x_1,x_2,
     L_T(z)+\left(x_3-\frac{L}{2}\right)\left(\frac{L_T(z)-L_B(z)}{L}\right)\Big) & \textrm{ if } x_3
     \in \left[-\frac{L}{2} -\delta, \frac{L}{2} +\delta\right]\\
     \big(x_1,x_2, x_3+L_T (z)\big) & \textrm{ if } x_3 \in \left[\frac{L}{2} + 2\delta, \infty\right) \\
     \big(x_1,x_2, x_3-L_B (z)\big) & \textrm{ if } x_3 \in \left(-\infty , -\frac{L}{2} - 2\delta\right] \\
     \big(x_1,x_2, \phi_1(x_3)\big) & \textrm{ if } x_3 \in \left[\frac{L}{2} + \delta, \frac{L}{2} + 2\delta\right] \\
     \big(x_1,x_2, \phi_2(x_3)\big) & \textrm{ if } x_3 \in \left[-\frac{L}{2} -2 \delta, -\frac{L}{2} -\delta\right], \end{array} \right.
\end{displaymath}
where  $\delta >0$ is a fixed number and $\phi_1, \phi_2$ are
smooth and one-to-one functions such that $\Phi$~is~smooth. Let us
notice that $\Phi$ for $x_3$ close to~facets is scaling in
direction $x_3$.

Now, we start to~construct the~map $\Psi$. The~idea comes from
differential geometry and~is~similar to~exponential map~$exp$ (see
also construction of Lie algebra $\mathfrak{g}$ of Lie group
$\textbf{G}$), we refer interested reader to~the textbook
\cite{Sternberg}. We construct a vector field $\Upsilon$ which
transforms the domain $\Sigma(z)$ onto $\Omega^c(z) $. Next we
integrate this vector field and we obtain the~flow. This flow gives
us~our~diffeomorphism.

Let us define the smooth map $\eta$
\begin{eqnarray*}
    \eta: \mathbb{R}^2 \rightarrow [0,1].
\end{eqnarray*}
We require that
\begin{displaymath}
    \eta (x_1,x_2)=\left\{ \begin{array}{ll} 1 & \textrm{ if } (x_1,x_2) \notin W(0) \textrm{ or } dist\big((x_1,x_2),\partial W(0)\big) < \frac{\varepsilon}{3}\\
                         0 & \textrm{ if } (x_1,x_2) \in W(0) \textrm{ and } dist\big((x_1,x_2),\partial W(0)\big) > \frac{2
                         \varepsilon}{3}.
               \end{array} \right.
\end{displaymath}
We will need it below.

 Subsequently, let us define the~map $\tilde{t}$ as
follows
\begin{eqnarray*}
    \tilde{t}(x_1,x_2)=\frac{\alpha_i'(x_1,x_2)}{\alpha'_{i}(x_1,x_2)+\alpha_i(x_1,x_2)},
\end{eqnarray*}
where $\alpha_i$ and $\alpha'_{i}$ are explained in Figure 4.
\begin{center}
\setcounter{figure}{3}
\begin{figure}
\centering
    \includegraphics[scale=.4]{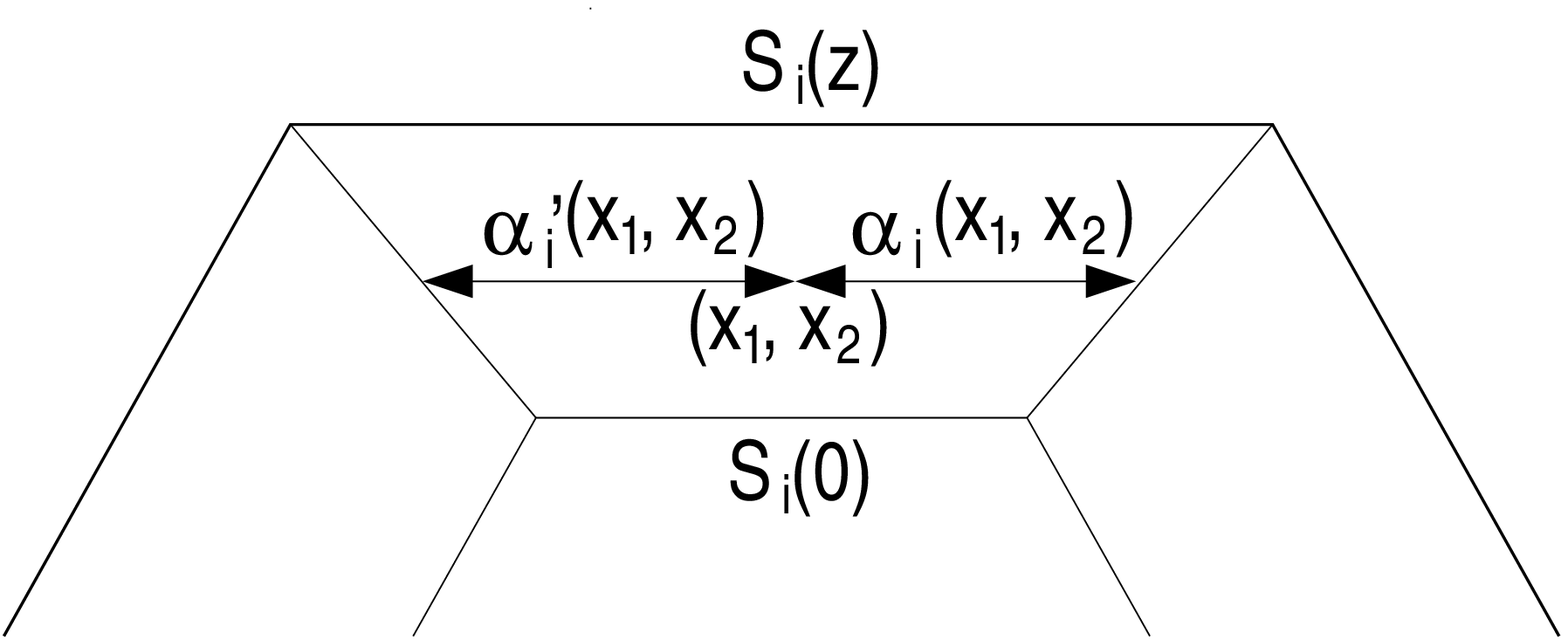}
\caption{}
\end{figure}
\end{center}
Next let us take a smooth map:
\begin{eqnarray*}
    \chi : [0,1] \rightarrow [0,1],
\end{eqnarray*}
such that $\chi _{[0,\frac{1}{3})}=0$, $\chi_{(\frac{2}{3},1]}=1$
and $\chi'(0)=\chi'(1)=0$. Let us define $\tau$ in the~following way
$\tau=\chi \circ \tilde{t}$. Let us notice that we can not take
simply $\tau =\tilde{t}$. Our vector field would have required the
properties without lines determined by directions $w_i$.

Subsequently, we define the~map $\xi_i$ as follows
\begin{eqnarray*}
    \xi_i=\big(1-\tau(x_1,x_2)\big)w_i+\tau(x_1,x_2)w_{i+1}.
\end{eqnarray*}

Our diffeomorphism must fulfil the~condition: the~boundary of
$N$-gon is transformed onto boundary of~$N$-gon. Now, we have to
construct the vector filed $\Upsilon$ in a smooth way. Namely,
\[
\Upsilon (x_1,x_2,x_3)=
\begin{cases}
\xi_i & \text{ if } (x_1,x_2) \in \Delta_i(z) \text{ and } x_3 \in
     \big(L_B(z),L_T(z)\big),\\
     \eta(x_1,x_2) \xi_i & \text{ if } (x_1,x_2) \in
     \Delta_i(z) \cup O_i(\varepsilon) \text{ and } x_3 \notin \big(L_B(t),L_T(z)\big),\\
     0 & \text{ in another cases}.
\end{cases}
\]
Where $\Delta_i(z)$ is $i$-th sector (see Figure 5)
\begin{figure}[h]
\centering
    \includegraphics[scale=.45]{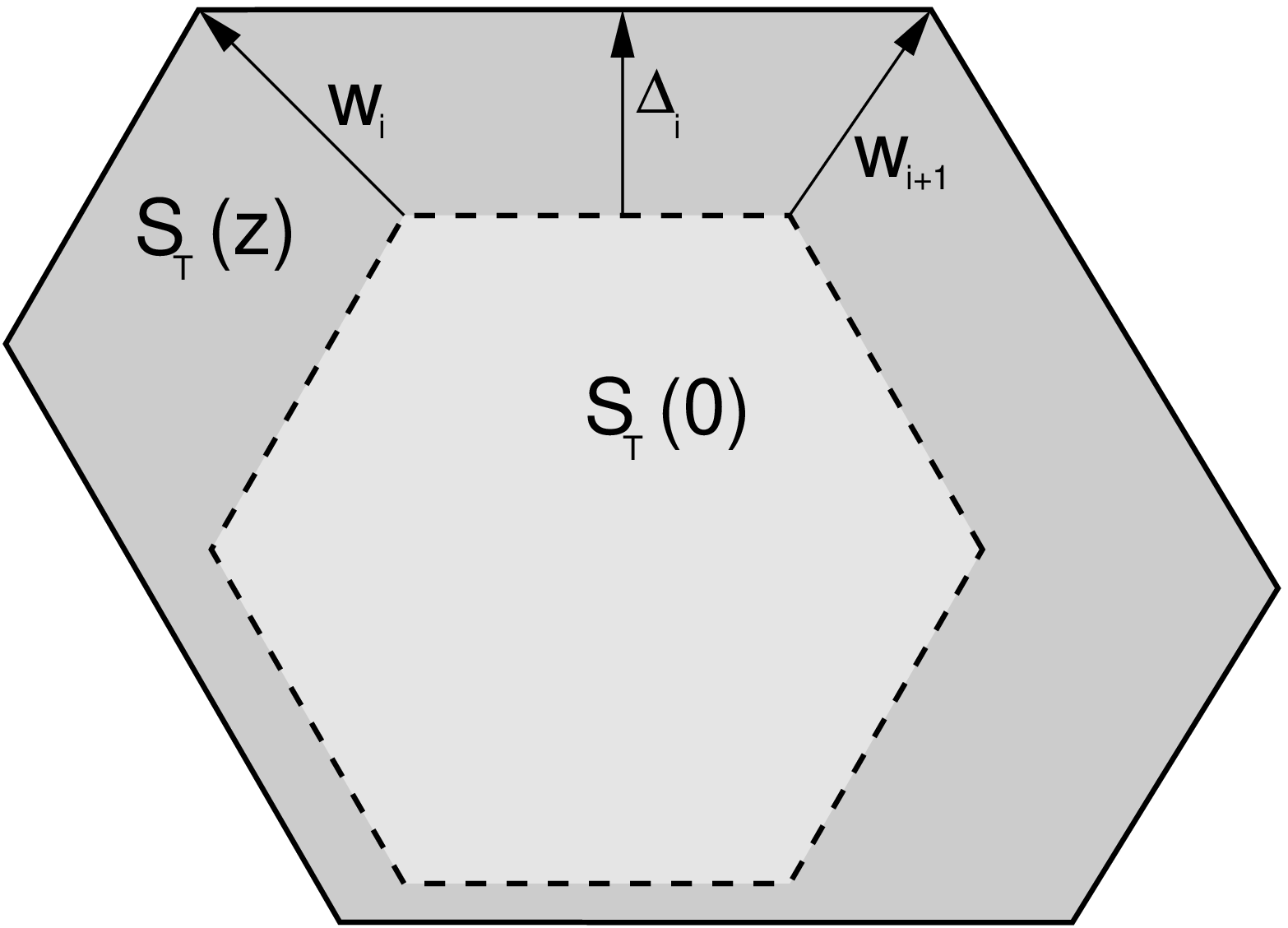}
\caption{}
\end{figure}
and
\begin{eqnarray*}
    O_i(\varepsilon)= (L_i(0))_{\varepsilon} \setminus( a_i(z) \cup
    b_i(z))
\end{eqnarray*}
where $(A)_{\varepsilon}$ is the $\varepsilon$ - envelope of set
$A$, i.e.
\begin{eqnarray*}
   (A)_{\varepsilon} = \{ x: dist(x, A) \leq \varepsilon \}.
\end{eqnarray*}

The sets $a_i(z) $ and $b_i(z)$ are explained in Figure 6.
\begin{figure}[h]
\centering
    \includegraphics[scale=.5]{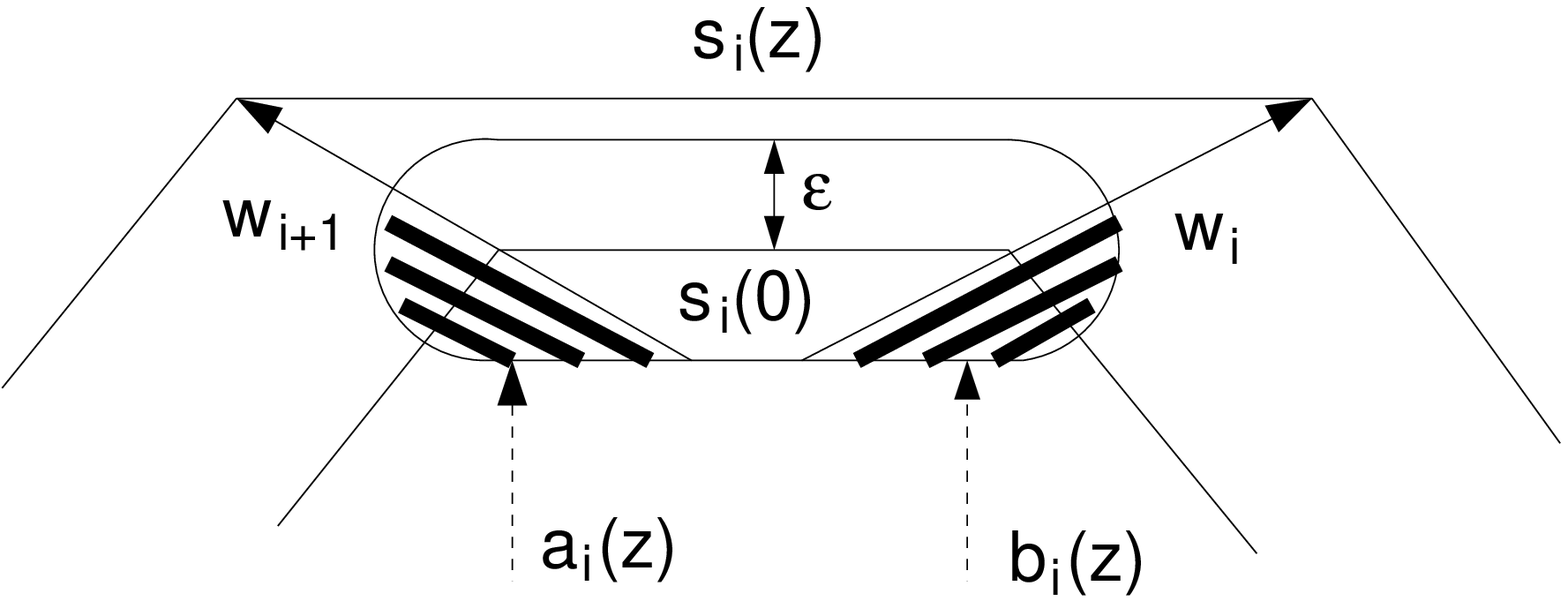}
\caption{}
\end{figure}
Thus we have obtained a smooth vector field. Integrating this
vector field we obtain the~flow $\Upsilon^t(x_1,x_2,x_3)$. We
define transformation $\Psi$ as follows
\begin{eqnarray*}
    \Psi (x_1,x_2,x_3)=\left( \begin{array}{cc} \Upsilon^1(x_1,x_2,x_3) \\ x_3 \end{array} \right).
\end{eqnarray*}
Finally our diffeomorphism has the~form:
\begin{eqnarray*}
    \Lambda (x_1,x_2,x_3) = \left( \begin{array}{cc} \Upsilon^1(x_1,x_2,x_3) \\ \Phi_3 (x_3)\end{array} \right) ,
\end{eqnarray*}
where $\Phi_3$ is the third component of $\Phi$.
\end{proof}
From the~proof of the above theorem we obtain the~following
\begin{wn}
    Let $J$ be the absolute value of Jacobian of transformation $\Lambda$, then:\\
    a) $J \in L_{\infty}$,\\
    b) if $|z_i| \leq M$, where $i=1,...,N+2$, then  there exist $l,L$ such that $0 < l \leq J \leq L$,\\
    c) $\Lambda$ is a~function of $(x_1,x_2,x_3)$ and $z_i(t)$.
\end{wn}
Now, we write the~system (\ref{problemu}) in the new variables. It
is not hard to~see that
\begin{eqnarray*}
    \nabla_{(\tilde{x_1},\tilde{x_2},\tilde{x_3})} = \left(\nabla_{(x_1,x_2,x_3)}\Lambda(x_1,x_2,x_3,\tau)\right)^{-1} \nabla_{(x_1,x_2,x_3)}\\
    \frac{\partial}{\partial t} = \frac{\partial}{\partial \tau}-\left(\frac{\partial \Lambda}{\partial
    \tau}\right)^{T}\left(\nabla_{(x_1,x_2,x_3)}\Lambda(x_1,x_2,x_3,\tau)\right)^{-1}\nabla_{(x_1,x_2,x_3)}\\
    \nu = \frac{\left(\nabla_{(x_1,x_2,x_3)}\Lambda(x_1,x_2,x_3,\tau)\right)^{-1}
    \bf{n}}{\left|\left(\nabla_{(x_1,x_2,x_3)}\Lambda(x_1,x_2,x_3,\tau)\right)^{-1} \bf{n}\right|}.
\end{eqnarray*}
We will use the~following notation:
\begin{eqnarray*}
    u(x_1,x_2,x_3,\tau):=u(\tilde{x_1},\tilde{x_2},\tilde{x_3},t)|_{(\tilde{x_1},\tilde{x_2},\tilde{x_3},t)=(\Lambda(x_1,x_2,x_3,\tau),\tau)}.
\end{eqnarray*}
Let us denote by $H(t)$ the transformed Laplace operator and
\begin{eqnarray*}
    \tilde{H}(t,z,\dot{z})\nabla_{(x_1,x_2,x_3)} =\left(\left(\frac{\partial \Lambda}{\partial
    \tau}\right)^{T} +
    f\right)\left(\nabla_{(x_1,x_2,x_3)}\Lambda(x_1,x_2,x_3,\tau)\right)^{-1}\nabla_{(x_1,x_2,x_3)}.
\end{eqnarray*}
The boundary operator is defined as follows
\begin{eqnarray*}
    B_0(t) \sigma= \frac{\left(\nabla_{(x_1,x_2,x_3)}\Lambda(x_1,x_2,x_3,\tau)\right)^{-1}
    \bf{n}}{\left|\left(\nabla_{(x_1,x_2,x_3)}\Lambda(x_1,x_2,x_3,\tau)\right)^{-1} \bf{n}\right|}\cdot \big(A(t),B(t),C(t)\big)
    \sigma
\end{eqnarray*}
where  $A,B$ and $C$ are defined in the following way

\begin{eqnarray*}
    A(\tau)=\frac{1}{Q} \left(\Lambda^2_{x_2} \Lambda^3_{x_3}\partial_{x_1}-\Lambda^2_{x_1}\Lambda^3_{x_3}\partial_{x_2}\right),\\
    B(\tau)=\frac{1}{Q} \left(\Lambda^1_{x_1}\Lambda^3_{x_3}\partial_{x_2}-\Lambda^1_{x_2}\Lambda^3_{x_3}\partial_{x_1}\right),\\
    C(\tau)=\frac{1}{Q} \left(\left(\Lambda^1_{x_2}\Lambda^2_{x_3}-\Lambda^1_{x_3}\Lambda^2_{x_2}\right)\partial_{x_1}+\left(\Lambda^1_{x_3}\Lambda^2_{x_1}-\Lambda^1_{x_1}\Lambda^2_{x_3}\right)\partial_{x_2}+\left(\Lambda^1_{x_1}\Lambda^2_{x_2}-\Lambda^1_{x_2}\Lambda^2_{x_1}\right)\partial_{x_3}\right),\\
    Q=\Lambda^3_{x_3} (\Lambda^1_{x_1} \Lambda^2_{x_2}-\Lambda^1_{x_2}\Lambda^2_{x_1}).
\end{eqnarray*}
We have used the notation $\Lambda^i_{x_j} =\frac{\partial \Lambda^i}{\partial x_j}$.
 Finally, system (\ref{problemu}) takes the~form
\begin{eqnarray}
    \frac{\partial}{\partial t} u=H\big(z(t)\big) u + \tilde{H}\big(z(t),\dot{z}(t)\big) \nabla u +
    \left(H(t)+\tilde{H}(t)\nabla
    -\frac{\partial}{\partial t} \right)G \text{ in } \Omega_0^c, \nonumber\\
    B_0(t)u=0  \text{  on  } \partial \Omega_0^c,\\
    \frac{1}{\mathcal{H}^2(S_i(t))} \int_{S_i(0)} \big(u(x) + G\big) J\big(x,z(t)\big)\,d\mathcal{H}^2(x) =\kappa_i(t)+\beta_i V_i
    ,\nonumber\\
    u(0)=u_0, \, z(0)=0. \nonumber
\end{eqnarray}
where $f,G,u$ are old $f,G,u$ in new variables.

\section {The existence of a solution - a linear theory}

In this section we show solvability of the~following problem
\begin{eqnarray*}
     \frac{\partial}{\partial t} u=H(t) u -\delta u + f(t) \text{ in } \Omega_0^c,\\
     B_0(t)u=0  \text{ on } \partial \Omega_0^c,\\
     u(0)=u_0,
\end{eqnarray*}
where $\delta$ will be specified later and $f$ is a H\"older continuous map.\\
Let us define an operator with the~domain
\begin{eqnarray*}
    H(t): D\big(H(t),\Omega^c\big) \rightarrow L^2(\Omega^c)\\
    D\big(H(t),\Omega^c\big):=\left\{ v\in H^1(\Omega^c): H(t)v \in L^2(\Omega^c), B_0(t) v|_{\partial \Omega^c} =0 \right\}.
\end{eqnarray*}
Where the~norm is defined as follows
\begin{eqnarray*}
    |\hspace{-0.5mm}|u|\hspace{-0.5mm}| _ {D(H(t),\Omega^c)}:= \left(|\hspace{-0.5mm}|u|\hspace{-0.5mm}|^2_ { H^1(\Omega^c)}+ |\hspace{-0.5mm}|H(t)u|\hspace{-0.5mm}|^2_ {  L^2(\Omega^c)}\right)^
    {\frac{1}{2}}.
\end{eqnarray*}
Later we will need the~following
\begin{prop}
    $D(H(t),\Omega^c)$ is Banach space for each $t \geq 0$.
\end{prop}
\begin{proof}
We know that $D(H(t),\Omega^c)$ and $D(\Delta,\Omega^c(t))$ are
isomorphic (see the~proof of Corollary \ref{izo}). Hence, it is
enough to~show that $D(\Delta,\Omega^c(t))$ is Banach space.

Let us take Cauchy sequence $\{u_k\}$ from
$D(\Delta,\Omega^c(t))$. Subsequently, $D_{x_i} u_k$ and $\Delta
u_k$ are Cauchy sequences in $L^2(\Omega^c(t))$. So $D_{x_i} u_k
\rightarrow v_i$ and $\Delta u_k \rightarrow f$ in
$L^2(\Omega^c(t))$. In particular $u_k \rightarrow u$
in~$L^2(\Omega^c(t))$. Hence,
\begin{eqnarray*}
    \int_{\Omega^c(t)} u D_{x_i} \phi dx \leftarrow \int_{\Omega^c(t)} u_k D_{x_i} \phi dx =(-1) \int_{\Omega^c(t)} D_{x_i}u_k \phi dx
    \rightarrow (-1) \int_{\Omega^c(t)} v_i \phi dx.
\end{eqnarray*}
Analogously
\begin{eqnarray*}
    \int_{\Omega^c(t)} u \Delta \phi dx \leftarrow \int_{\Omega^c(t)} u_k \Delta \phi dx =(-1)^2 \int_{\Omega^c(t)} \Delta u_k \phi dx
    \rightarrow (-1)^2 \int_{\Omega^c(t)} f \phi dx.
\end{eqnarray*}
Hence, we obtain $D_{x_i}u=v_i$ and $\Delta u =f$. From trace
theorem and $D(\Delta,\Omega^c(t)) \subset H^{\frac{3}{2}
+\varepsilon}(\Omega^c(t))$ (see \cite{Grisvard}) we obtain
\begin{eqnarray*}
    \left|\hspace{-0.5mm}\left|\frac { \partial (u_l -u_k)} {\partial \nu}\right|\hspace{-0.5mm}\right|_ {L^{2} ( \partial \Omega^c(t))} \leq
    C |\hspace{-0.5mm}|u_l -u_k|\hspace{-0.5mm}|_ {D(\Delta,\Omega^c (t))}.
\end{eqnarray*}
Hence, we obtain the~assertion of the~proposition.
\end{proof}
Now, we show the~following
\begin{lem}
\label{podciag} If $u \in D(\Delta,\Omega)$ then there exists
a~sequence $\{R_n \}$,  $R_n \rightarrow \infty$ such that
\begin{eqnarray*}
     \lim_{R_n \rightarrow \infty} \int_{\partial B(0,R_n)} \frac{\partial
    u}{\partial \nu} \bar{u} \,d \mathcal{H}^2 = 0
\end {eqnarray*}
\end{lem}
\begin{proof}
From the integration by parts formula the~following expression
follows
\begin{eqnarray*}
    \int_ {\Omega^c \setminus B(0,R)} u \Delta u dx + \int_ {\Omega^c \setminus B(0,R)} |\nabla u|^2 dx = \int_ {\partial B(0,R)} \frac{\partial
    u}{\partial \nu} \bar{u} \,d \mathcal{H}^2.
\end{eqnarray*}
But $u, \nabla u, \Delta u \in L^2(\Omega^c)$. Hence there exists
a~sequence $R_n$ such that
\begin{eqnarray*}
    \int_ {\Omega^c \setminus B(0,R_n)} u \Delta u dx \rightarrow 0,\\
     \int_ {\Omega^c \setminus B(0,R_n)} |\nabla u|^2 dx \rightarrow 0
\end{eqnarray*}
as $R_n \rightarrow \infty$.\\
From this the~lemma follows.
\end{proof}
Now, we can go to the following result.

\begin{twr}
\label{gen}
Each of the~operators  $-H(t)$, $t \geq 0$, generates an analytic semigroup.
\end{twr}
\begin{proof}
It is enough to~show that $-H(t)$ is a sectorial operator (see
\cite{Lunardi} or \cite{Henry}). First of all we~examine
the~operator
 $H(0)$, i.e. $\Delta$. Namely, we~show the~following,
\begin{lem}
\label{deltagen}
The operator $-\Delta$ with the~domain $D(\Delta, \Omega^c)$ is sectorial in $L^2(\Omega ^c)$.
\end{lem}
\begin{proof}
First of all we show that the operator $-\Delta$ is closed and
densely defined. Indeed, since $C^{\infty}_0(\Omega) \subset
D(\Delta, \Omega)$ and $C^{\infty}_0(\Omega)$ is dense in
$L^2(\Omega)$, hence $D(\Delta, \Omega)$ is dense in
$L^2(\Omega)$. Now,~we~show that~$-\Delta$ is a~closed operator.

We need the~following estimate
\begin{eqnarray}
    \label{oszacowanie}
    |\hspace{-0.5mm}|\nabla v|\hspace{-0.5mm}|^2_ {L^2(\Omega^c)} \leq |\hspace{-0.5mm}|v|\hspace{-0.5mm}|_ {L^2(\Omega^c)} |\hspace{-0.5mm}|\Delta v|\hspace{-0.5mm}|_ {L^2(\Omega^c)}
\end{eqnarray}
for $v \in D(\Delta , \Omega^c)$. In order to~show inequality
(\ref{oszacowanie}) we take the~following equality
\begin{eqnarray*}
    \Delta v =f
\end{eqnarray*}
where $f \in L^2(\Omega^c)$ and $v \in D(\Delta)$. We multiply
both sides of the above equality by $v$ and after integration by
parts we get,
\begin{eqnarray*}
    |\hspace{-0.5mm}|\nabla v|\hspace{-0.5mm}|^2_ {L^2(\Omega^c)} = \int_ {\Omega^c} v f dx \leq  |\hspace{-0.5mm}|v|\hspace{-0.5mm}|_ {L^2(\Omega^c)} |\hspace{-0.5mm}|f|\hspace{-0.5mm}|_ {L^2(\Omega^c)}
     = |\hspace{-0.5mm}|v|\hspace{-0.5mm}|_ {L^2(\Omega^c)} |\hspace{-0.5mm}|\Delta v|\hspace{-0.5mm}|_ {L^2(\Omega^c)},
\end{eqnarray*}
where we used Lemma \ref{podciag} and Schwarz inequality.\\

Now, let us take a~sequence $\{v_k \}_{k=1}^{\infty} \subset
D(\Delta, \Omega^c)$ such that:
\begin{eqnarray*}
    v_k \rightarrow v\\
    \Delta v_k \rightarrow f
\end{eqnarray*}
in $L^2(\Omega^c)$. From inequality (\ref{oszacowanie}) we obtain
\begin{eqnarray*}
    |\hspace{-0.5mm}|v_k - v_l|\hspace{-0.5mm}|^2_ {D (\Delta, \Omega^c)} =|\hspace{-0.5mm}|v_k-v_l|\hspace{-0.5mm}|^2_ { H^1 (\Omega^c)}+ |\hspace{-0.5mm}|\Delta v_k - \Delta v_l|\hspace{-0.5mm}|^2_ { L^2(\Omega^c)} = \\
    |\hspace{-0.5mm}|v_k-v_l|\hspace{-0.5mm}|^2_ { L^2(\Omega^c)}+ |\hspace{-0.5mm}|\nabla v_k-\nabla v_l|\hspace{-0.5mm}|^2_ { L^2(\Omega^c)}+|\hspace{-0.5mm}|\Delta v_k - \Delta v_l|\hspace{-0.5mm}|^2_ { L^2(\Omega^c)} \leq \\
    |\hspace{-0.5mm}|v_k-v_l|\hspace{-0.5mm}|^2_ { L^2(\Omega^c)}+ |\hspace{-0.5mm}| v_k- v_l|\hspace{-0.5mm}|_ { L^2 (\Omega^c)} |\hspace{-0.5mm}|\Delta v_k - \Delta v_l|\hspace{-0.5mm}|_ { L^2(\Omega^c)} +|\hspace{-0.5mm}|\Delta v_k - \Delta
    v_l|\hspace{-0.5mm}|^2_ { L^2(\Omega^c)} \leq \\\left(|\hspace{-0.5mm}| v_k- v_l|\hspace{-0.5mm}|_ { L^2(\Omega^c)} +|\hspace{-0.5mm}|\Delta v_k - \Delta v_l|\hspace{-0.5mm}|_ { L^2(\Omega^c)}\right)^2.
\end{eqnarray*}
Hence
\begin{eqnarray*}
    |\hspace{-0.5mm}|v_k - v_l|\hspace{-0.5mm}|_ {D(\Delta, \Omega^c)} \leq |\hspace{-0.5mm}| v_k- v_l|\hspace{-0.5mm}|_ { L^2(\Omega^c)} +|\hspace{-0.5mm}|\Delta v_k - \Delta v_l|\hspace{-0.5mm}|_ { L^2(\Omega^c)}.
\end{eqnarray*}
Subsequently, we obtain that  $\{v_k \}_{k=1}^{\infty}$ is Cauchy
sequence in $D(\Delta, \Omega^c)$. Finally,
\begin{eqnarray*}
    v_k \rightarrow v \text{  w   } D(\Delta, \Omega^c).
\end{eqnarray*}
Hence, $v\in D(\Delta, \Omega^c)$ and $\Delta v_k \rightarrow
\Delta v$.

Now, we show the estimate for a resolvent operator. We follow the
method used in the book \cite{Henry}. Let us consider the~equation
\begin{eqnarray*}
\lambda u +\Delta u =f
\end{eqnarray*}
where $f \in L^2$ and $u \in D(\Delta)$. Let us assume that $Re \lambda < 0$.\\
Let us multiply the above equation by $\bar{u}$ and integrate:
\begin{eqnarray*}
    \int_{\Omega^c}\lambda |u|^2 +\bar{u} \Delta u =\int_{\Omega^c} \bar{u} f.
\end{eqnarray*}
Integrating by parts we obtain
\begin{eqnarray*}
    \int_{\Omega^c}\lambda |u|^2 -|\nabla u|^2 =\int_{\Omega^c} \bar{u} f,
\end{eqnarray*}
where we applied Lemma \ref{podciag}.

Next, we take the real and imaginary part of above expression:
\begin{eqnarray*}
    Re \lambda |\hspace{-0.5mm}|u|\hspace{-0.5mm}|^2_{L^2} - |\hspace{-0.5mm}| \nabla u |\hspace{-0.5mm}|^2 _{L^2} = Re \int_ {\Omega^c} \bar{u} f\\
    -Im \lambda |\hspace{-0.5mm}|u|\hspace{-0.5mm}|^2_{L^2} = Im \int_ {\Omega^c} \bar{u} f.
\end{eqnarray*}
Hence,
\begin{eqnarray*}
    Re \lambda |\hspace{-0.5mm}|u|\hspace{-0.5mm}|_{L^2(\Omega^c)}  \le |\hspace{-0.5mm}|f|\hspace{-0.5mm}|_{L^2(\Omega^c)} \\
    |Im \lambda |\,|\hspace{-0.5mm}|u|\hspace{-0.5mm}|_{L^2(\Omega^c)} \le |\hspace{-0.5mm}|f|\hspace{-0.5mm}|_{L^2(\Omega^c)}.
\end{eqnarray*}
Subsequently, we take a~sum of the above inequalities
\begin{eqnarray*}
    |\lambda |\,|\hspace{-0.5mm}|u|\hspace{-0.5mm}|_{L^2(\Omega^c)} \le 2 |\hspace{-0.5mm}|f|\hspace{-0.5mm}|_{L^2(\Omega^c)}.
\end{eqnarray*}
Next, for arbitrary $\lambda$ we behave in a standard way. It
means, if $Re \lambda >0$ and $\phi \leq |\arg(\lambda)| \leq
\pi$, let us denote $\lambda =a + i b$ we obtain $\tan \phi \leq
\left|\frac{b}{a}\right|$. From the above estimates we obtain:
\begin{eqnarray*}
    |b|\,|\hspace{-0.5mm}|u|\hspace{-0.5mm}|_{L^2(\Omega^c)} \le |\hspace{-0.5mm}|f|\hspace{-0.5mm}|_{L^2(\Omega^c)}
\end{eqnarray*}
and
\begin{eqnarray*}
    |a|\,|\hspace{-0.5mm}|u|\hspace{-0.5mm}|_{L^2} \le \frac{|\hspace{-0.5mm}|f|\hspace{-0.5mm}|_{L^2}}{tg \phi}.
\end{eqnarray*}
Hence,
\begin{eqnarray*}
    |\lambda |\,|\hspace{-0.5mm}|u|\hspace{-0.5mm}|_{L^2(\Omega^c)} \le \frac{1}{|\sin \phi|} |\hspace{-0.5mm}|f|\hspace{-0.5mm}|_{L^2(\Omega^c)}.
\end{eqnarray*}
This ends the~proof of the~lemma.
\end{proof}
Now, we show Theorem \ref{gen} for arbitrary $t>0$.

Using the~same argument as proof of Lemma \ref{deltagen} one can
show $-H(t)$ is densely defined. Subsequently, we show that
the~operator $-H(t)$ is closed. First of all we show the~following
inequalities
\begin{eqnarray}
\label{niero}
    |\hspace{-0.5mm}|\nabla u|\hspace{-0.5mm}|^2_ {L^2(\Omega^c)} \leq c|\hspace{-0.5mm}|H(t)u|\hspace{-0.5mm}|_
    {L^2(\Omega^c)}\,|\hspace{-0.5mm}|u|\hspace{-0.5mm}|_ {L^2(\Omega^c)} ,
\end{eqnarray}
where $u \in D(H(t),\Omega^c)$.
Indeed, let us consider the~equation
\begin{eqnarray*}
    H(t)u =f \text{ in } \Omega^c.
\end{eqnarray*}
Changing the~variables in the~above equation we obtain the~following one
\begin{eqnarray*}
    \Delta u =f \text { in } \Omega^c(t).
\end{eqnarray*}
where $u$ and $f$ are old $u$ and $f$ in the~new variables.
Following  the~method presented in the~proof of Lemma
\ref{deltagen} and from Corollary 1 we obtain the~inequality
(\ref{niero}).

From inequality (\ref{niero}) closedness easily follows. Indeed,
let us take any sequence $\{u_k\} \subset D(H(t), \Omega)$ such
that $u_k \rightarrow u $ and $H(t)u_k \rightarrow f$ in
$L^2(\Omega)$. Hence $u_k$ and $H(t)u_k$ are Cauchy sequences in
$L^2(\Omega)$. Subsequently, from inequality (\ref{niero}) we
obtain the~following estimate
\begin{eqnarray*}
    |\hspace{-0.5mm}|\nabla u_k -\nabla u_l|\hspace{-0.5mm}|^2_{L^2(\Omega^c)} \leq c|\hspace{-0.5mm}|H(t)(u_k - u_l)|\hspace{-0.5mm}|_{L^2(\Omega^c)}
     \,|\hspace{-0.5mm}|u_k - u_l|\hspace{-0.5mm}|_{L^2(\Omega^c)}
\end{eqnarray*}
and this inequality finishes the closedness of the~operator $H(t)$
(see proof the~Lemma \ref{deltagen} for details).

Now, we show the estimate for the resolvent operator. Let us
consider the~equation
\begin{eqnarray*}
    \lambda u + H(t) u = f  \text{ in } \Omega^c.
\end{eqnarray*}
where $f \in L^2$ and $u \in D(H(t),\Omega^c)$. Changing
the~variables in the~above equation we obtain the~following
equation
\begin{eqnarray*}
    \lambda u + \Delta u = f  \text{ in } \Omega^c (t),
\end{eqnarray*}
where $u$ and $f$ are old $u$ and $f$ in the~new variables.

From the proof of the~Lemma \ref{deltagen} we obtain
the~inequality
\begin{eqnarray*}
    |\lambda |\,|\hspace{-0.5mm}|u|\hspace{-0.5mm}|_{L^2(\Omega^c_t)} \le C |\hspace{-0.5mm}|f|\hspace{-0.5mm}|_{L^2(\Omega^c_t)}.
\end{eqnarray*}
After changing the~variables we obtain
\begin{eqnarray*}
|\lambda| \int_{\Omega^c} |u|^2 J dx \le C \int_{\Omega^c}|f|^2 J dx.
\end{eqnarray*}
From Corollary 1 the following
\begin{eqnarray*}
    |\lambda |\,|\hspace{-0.5mm}|u|\hspace{-0.5mm}|_{L^2(\Omega^c)} \le C \frac{L}{l}|\hspace{-0.5mm}|f|\hspace{-0.5mm}|_{L^2(\Omega^c)}
\end{eqnarray*}
holds.

 From this the~theorem follows.
\end{proof}

Before we state next result we recall the definition of some
hypothesis (see \cite{Ac} and \cite{Ac1}). Namely,\\

\textbf{Hypothesis I}. For each $t \in [0,T]$, $A(t) :D(A(t))
\subseteq L^2 \rightarrow L^2$ is a closed linear operator and
there exist $M>0$ and $\theta_0 \in (\frac{\pi}{2},\pi]$ such that
\begin{eqnarray*}
   \rho(A(t))  \supseteq S_{\theta_0} = \{ z \in \mathbb{C} : |\arg z| \leq \theta_0 \} \text{   } \forall t \in [0,T],\\
   ||R(\lambda,A(t))||_{{\cal{L}}(L^2)} \leq \frac{M}{1+|\lambda|}
   \text{  } \forall \lambda \in S_{\theta_0}, \forall t \in
   [0,T].
\end{eqnarray*}
\textbf{Hypothesis II}. There exist $B >0$, $ k \in \mathbb{N}$,
$\alpha_1 ,..., \alpha_k , \beta_1,...,\beta_k$ with $0 \leq
\beta_i < \alpha_i \leq 2$, such that
\begin{eqnarray*}
   ||A(t)R(\lambda,A(t))[A(t)^{-1} - A(s)^{-1}]||_{{\cal{L}}(L^2)} \leq B \sum_{i=1}^k (t-s)^{\alpha_i} |\lambda|^{\beta_i -1} \text{  }
\end{eqnarray*}
$\forall \lambda \in S_{\theta_0}\setminus \{0\}$, $\forall 0\leq
s \leq t \leq T$. Where $R(\lambda,A(t))$ is resolvent operator.

 We have the~following
\begin{twr}
There exists $\delta >0$ such that the~operator
\begin{eqnarray*}
    A(t) = H(t) - \delta Id
\end{eqnarray*}
with a~domain $D(A(t))=D(H(t))$ fulfills Hypothesis I and
Hypothesis II.
\end{twr}
\begin{proof}
The proof of Hypothesis I is almost the~same as the proof of
Theorem \ref{gen}.

Concerning Hypothesis II, if we set for$f \in L^2(\Omega^c)$
\begin{eqnarray*}
    v= A(s)^{-1} f, \text{   }  u = R\big(\lambda ,A(t)\big)\big[\lambda - A(s)\big] v,
\end{eqnarray*}
then we have to~estimate $L^2$-norm of
\begin{eqnarray*}
    u- v= A(t)R\big(\lambda ,A(t)\big)\left[A(t)^{-1} - A(s)^{-1}\right]f.
\end{eqnarray*}
Now, $u$ and $v$ solve respectively,
\begin{eqnarray*}
    \lambda u -A(t) u = \lambda v -f \text{  in  } \Omega^c\\
    B(t)u = 0 \text{  on  } \partial \Omega^c
\end{eqnarray*}
and
\begin{eqnarray*}
    A(s)v = -f \text{  in  } \Omega^c\\
    B(s)v = 0 \text{  on  } \partial \Omega^c.
\end{eqnarray*}
Hence, $u-v$ solves
\begin{eqnarray*}
    \lambda (u-v) + A(s) v -A(t) u = 0 \text{  in  } \Omega^c\\
    B(t)u = B(s)v \text{  on  } \partial \Omega^c.
\end{eqnarray*}
Next, using the~similar methods as in the~proof of Theorem \ref{gen}
and from inequality
\begin{eqnarray*}
    |\hspace{-0.5mm}|\nabla v|\hspace{-0.5mm}|_{L^2(\Omega^c)} \leq c|\hspace{-0.5mm}|f|\hspace{-0.5mm}|_{L^2(\Omega^c)}
\end{eqnarray*}
and Lipschitz continuity of coefficients of the operator $A$ the
proof follows. 
\end{proof}
Now, from the above theorem and \cite{Ac} the~following holds.
\begin{wn}
If $f$ is a H\"older continuous map, then there exists unique
solution of the~following problem
\begin{eqnarray*}
     \frac{\partial}{\partial t} u=A(t) u + f(t) \text{  on  } \Omega_0^c,\\
     B_0(t)u=0  \text{  in  } \partial \Omega_0^c,\\
     u(0)=u_0.
\end{eqnarray*}
It is given by the formula
\begin{eqnarray*}
    u (t) = U(t,0)u_0 +\int_{0}^t  U(t,\sigma)f(\sigma) d \sigma.
\end{eqnarray*}
\end{wn}

\section{Nonlinear theory}

Before we state the~main result we are going to present some facts.

 In order to understand better interpolation spaces we show
 isomorphism theorem.
\begin{wn}
     \label{izo}

    The following interpolation spaces are isomorphic
    \begin{eqnarray*}
            \left[L^2 (\Omega^c _t), D(\Delta, \Omega^c _t)\right]_{\theta} \backsimeq \left[L^2(\Omega^c ), D(H(t), \Omega^c)\right]_{\theta}.
    \end{eqnarray*}
\end{wn}
\begin{proof}
From Theorem \ref{dif} we know that there is a family of
$C^{2+\alpha}$ diffeomorphisms $\Lambda_t :\Omega^c _t \rightarrow
\Omega^c$. Let us denote by $\Lambda _{t_{*}}$ the pullback
operator. Hence we can write the~following diagram
$$\xymatrix{& \Omega^c_t \ar[r]^{\Lambda_t } \ar[ld] & \Omega^c \ar[dr] & \\
    L^2(\Omega^c_t)  \ar[rd] & & & L^2(\Omega^c) \ar[ld]
    \ar[lll]_{\Lambda_{t_{*}}}^
    {\backsimeq} \\
    & (L^2(\Omega^c_t),D(\Delta,\Omega^c_t))  &
    (L^2(\Omega^c),D(H(t),\Omega^c)) \ar[l]_{\backsimeq} &
    \\
    D(\Delta,\Omega^c_t) \ar[ru] & & & D(H(t),\Omega^c) \ar[lll]_{\Lambda_{t_{*}}}^
    {\backsimeq} \ar[ul] \\
    & \Omega^c_t \ar[r]^{\Lambda_t } \ar[lu] & \Omega^c \ar[ru] &\\
 }$$

So we obtain an isomorphism on the~level of interpolation pairs.
It is well known (see \cite{Triebel}) that interpolation is a
functor from a category of Banach pairs into a category of
interpolation spaces. Hence, from functoriality
(see~\cite{Spanier} or~\cite{Maclane}) we obtain
$$\xymatrix{ (L^2(\Omega^c_t),D(\Delta,\Omega^c_t)) \ar[r]_{\backsimeq} \ar[d]^{\theta}  &
    (L^2(\Omega^c),D(H(t),\Omega^c))\ar[d]^{\theta}  \\
    [L^2(\Omega^c_t),D(\Delta,\Omega^c_t))]_{\theta} \ar[r]_{\backsimeq} &
    [(L^2(\Omega^c),D(H(t),\Omega^c))]_{\theta}   \\
 }$$

Where $\theta$ is the interpolation functor.
\end{proof}
\begin{wn}
\label{zan}
 If $\beta > \frac{s}{\frac{3}{2} + \epsilon}$, $\epsilon$ is sufficiently small and $s < \frac{3}{2} +\epsilon$, then the following is
 true
\begin{eqnarray*}
    \left[L^2(\Omega^c ), D(H(t), \Omega^c)\right]_{\beta} \hookrightarrow H^s
(\Omega^c).
\end{eqnarray*}
\end{wn}
\begin{proof}
It follows immediately from the properties of interpolation spaces
and Grisvard theory (see~\cite{Grisvard} and~\cite{Grisvard1}),
see also \cite{Triebel}.
\end{proof}

Now, we show a compactness result.
\begin{lem}
\label{zwartosc} Let us assume $\Omega \subset \mathbb{R}^k$ is
bounded domain with Lipschitz boundary. Let us also assume that
the sequence ${u_n}$ is bounded in $H^{\sigma}(\mathbb{R}^k
\setminus \Omega)$ and $||u_n||_{H^{\sigma}(\mathbb{R}^k \setminus
B(0,N))} \rightarrow 0$ as $N \rightarrow \infty$ uniformly with
respect to~$n$. Then there exists subsequence of $u_n$ convergence
in $H^{\alpha}(\mathbb{R}^k \setminus \Omega)$ for $\alpha <
\sigma$.
\end{lem}
\begin{proof}
We can assume that $\Omega \subset B(0,1)$. Let $\{u_n\} \subset
H^{\sigma}(\mathbb{R}^k \setminus \Omega)$, then $\{u_n
|_{B(0,N)}\} \subset H^{\sigma}(B(0,N) \setminus \Omega)$. Using
Rellich - Kondrachov Theorem we can extract a~sequence
$\{u_{n_{1}} |_{B(0,1)}\}\subset H^{\sigma}(B(0,1) \cap \Omega)$
such that  $\{u_{n_{1}} |_{B(0,1)}\} \rightarrow u_1$ in
$H^{\alpha}(B(0,1) \setminus \Omega)$. Let us denote by
$\tilde{u}_1$ extension of $u_1$ onto $\mathbb{R}^k \setminus
\Omega$. So we have chosen a~subsequence $\{u_{n_{1}} \}$ (we
restrict the sequence to~such that the~following inequality is
satisfied
$|\hspace{-0.5mm}|u_{n_{1}}-\tilde{u}_1|\hspace{-0.5mm}|_{H^{\alpha}(B(0,1)
\setminus \Omega)} \leq 1$). Analogously we work on ball $B(0,2)$,
e.g. we subtract subsequence $\{u_{n_{1}} |_{B(0,2)} \}$, (we will
denote it by $\{u_{n_{2}} |_{B(0,2)} \}$) such that $\{u_{n_{2}}
|_{B(0,2)}\} \rightarrow u_2$. Denote extension by $\tilde{u}_2$
and we take only this terms such
$|\hspace{-0.5mm}|u_{n_{2}}-\tilde{u}_2|\hspace{-0.5mm}|_{H^{\alpha}(B(0,2)
\setminus \Omega)} \leq \frac{1}{2}$ end so on, namely\\
$u_{n_{1}} \rightarrow u_1$ in $H^{\alpha}(B(0,1) \setminus
\Omega)$ and extension $\tilde{u}_1$ \\ and such that
$|\hspace{-0.5mm}|u_{n_{1}}-\tilde{u}_1|\hspace{-0.5mm}|_{H^{\alpha}(B(0,1)
\setminus \Omega)} \leq 1$\\
$u_{n_{2}} \rightarrow u_2$ in $H^{\alpha}(B(0,2) \setminus
\Omega)$ and extension $\tilde{u}_2$ \\ and such that
$|\hspace{-0.5mm}|u_{n_{2}}-\tilde{u}_2|\hspace{-0.5mm}|_{H^{\alpha}(B(0,2)
\setminus \Omega)} \leq \frac{1}{2}$\\
$u_{n_{3}} \rightarrow u_3$ in $H^{\alpha}(B(0,3) \setminus
\Omega)$ and extension $\tilde{u}_3$ \\ and such that
$|\hspace{-0.5mm}|u_{n_{3}}-\tilde{u}_3|\hspace{-0.5mm}|_{H^{\alpha}(B(0,3)
\setminus \Omega)} \leq \frac{1}{3}$\\
.\\
.\\
.\\
$u_{n_{N}} \rightarrow u_N$ in $H^{\alpha}(B(0,N) \setminus
\Omega)$ and extension $\tilde{u}_N$ \\ and such that
$|\hspace{-0.5mm}|u_{n_{N}}-u_N|\hspace{-0.5mm}|_{H^{\alpha}(B(0,N)
\setminus \Omega)} \leq \frac{1}{N}$\\
where $u_{n_{N}}$ is subsequence of $u_{n_{M}}$ for $N>M$. We
shall show that $\{ \tilde{u}_N \}$ is convergent in
$H^{\alpha}(\mathbb{R}^k \setminus \Omega)$. Indeed, we show that
Cauchy condition is fulfilled
\begin{eqnarray*}
    |\hspace{-0.5mm}|\tilde{u}_N-\tilde{u}_M|\hspace{-0.5mm}|_ {H^{\alpha} (\mathbb{R}^k \setminus \Omega)}
    \leq |\hspace{-0.5mm}|\tilde{u}_N-\tilde{u}_M|\hspace{-0.5mm}| _ {H^{\alpha} (\Omega \cap B(0, min
    \{N,M \})}+\\
    |\hspace{-0.5mm}|\tilde{u}_N|\hspace{-0.5mm}| _ {H^ {\alpha}
    (\mathbb{R}^k \setminus B(0, min \{N,M \}))} +
     |\hspace{-0.5mm}|\tilde{u}_M|\hspace{-0.5mm}|_ {H^ { \alpha}(\mathbb{R}^k \setminus B(0, min \{N,M
    \}))}.
\end{eqnarray*}
Hence, we obtain
\begin{eqnarray*}
    \underset{N,M \rightarrow
    \infty}{\lim \sup}|\hspace{-0.5mm}|\tilde{u}_N-
    \tilde{u}_M|\hspace{-0.5mm}|_{H^{\alpha}(\mathbb{R}^k \setminus \Omega)} =0.
\end{eqnarray*}
Let us denote the~limit of this sequence by $u$. We show that
$u_{n_{N}}$ converges to~$u$ in $H^{\alpha}(\mathbb{R}^k \setminus
\Omega)$. Namely,
\begin{eqnarray*}
    |\hspace{-0.5mm}|u_{n_{N}}-u |\hspace{-0.5mm}|_ { H^ { \alpha}(\mathbb{R}^k \setminus \Omega)}
    \leq |\hspace{-0.5mm}|u_{n_{N}}-u |\hspace{-0.5mm}|_ {H^ { \alpha}( B(0,N)\setminus\Omega )}+\\
    |\hspace{-0.5mm}|u_{n_{N}} |\hspace{-0.5mm}|_ { H^ {\alpha}( \mathbb{R}^k \setminus B(0, N))} +
     |\hspace{-0.5mm}|u|\hspace{-0.5mm}|_ {H^ {\alpha}(\mathbb{R}^k \setminus B(0,
    N))}.
\end{eqnarray*}
These last two terms converge to~$0$, and
\begin{eqnarray*}
    |\hspace{-0.5mm}|u_{n_{N}}-u |\hspace{-0.5mm}|_ {H^ {\alpha}(B(0,N) \setminus \Omega )}\leq |\hspace{-0.5mm}|u_{n_{N}}
    -\tilde{u}_N|\hspace{-0.5mm}|_ {H^ {\alpha}(B(0,N) \setminus \Omega )} + ||
    \tilde{u}_{N}-u |\hspace{-0.5mm}|_ {H^{\alpha}(B(0,N)\setminus \Omega )}\\
    \leq \frac{1}{N} + |\hspace{-0.5mm}|\tilde{u}_{N}-u |\hspace{-0.5mm}|_ {H^ {\alpha}(\mathbb{R}^3 \setminus \Omega
    )}.
\end{eqnarray*}
This finishes the~proof.
\end{proof}
Now, we can formulate the~main result.
\begin{twr}
Let us assume $u_0  \in D(A(0))$ and $g \in C^{1,\eta}((0,T);
\mathbb{R}^{N+2})$, then there exists a solution of the~problem
\begin{eqnarray*}
    \frac{\partial u}{\partial t} = H\big(z(t)\big)u+ \tilde{H}\big(z(t), \dot{z}(t)\big)\nabla
    u+\left(H(t)+\tilde{H}(t)\nabla
    -\frac{\partial}{\partial t} \right)G,\\
    B_0(t)u=0,\\
    \dot{z}_i=-\frac{1}{\beta_i} \kappa_i\big(z(t)\big)+ \frac{1}{\beta_i\mathcal{H}^2(S_i(z))}\int_{S_i(0)} \big(u(x)
    +G\big) J\big(x,z(t)\big) d\mathcal{H}^2 (x),\\
    u(0)=u_0, z(0)=0.
\end{eqnarray*}
such that $u \in C^{\alpha}([0,T);H^s(\Omega^c))$ and $z \in
(C^{1, \alpha}([0,T)))^{N+2}$. Where $1< s <\frac{3}{2} +\epsilon
$ and $\alpha <min (1-(\frac{3}{2}+\epsilon)^{-1}s, \eta)$.
\end{twr}
For the sake of simplicity of notation we shall write $L^2
=L^2(\Omega^c)$ and $H^s =H^s (\Omega^c)$.
\begin{proof}
We will apply the~method of successive approximation. We set
\begin{eqnarray*}
    u^{0}(t)=u_0, \text{      } t \in [0,\infty)
\end{eqnarray*}
and let $z_i^0$ be a~solution of the~following problem
\begin{eqnarray*}
    \frac{d}{d t}{z_i}^0= \frac{1}{\beta_i} \kappa_i\big(z^0(t)\big)+
    \frac{1}{\beta_i\mathcal{H}^2(S_i(z))}\int_{S_i(0)} \big(u^0(x) +G\big)J\big(x,z(t)\big) d\mathcal{H}^2 (x), \\
     z_i^0=0,
\end{eqnarray*}
on the~maximal interval of existence $[0,T_0)$. Next, we define
$u^{n+1}$  as an unique solution (on interval ~$[0,T_n)$)
of~the~following problem
\begin{eqnarray}
\label{par}
    \frac{\partial u^{n+1}}{\partial t} = H(t)u^{n+1}+ \tilde{H}(t)\nabla u^{n+1} +
     \left(H(t)+\tilde{H}(t)\nabla -\frac{\partial}{\partial t}
    \right)G \nonumber\\
    B_0 u^{n+1}=0\\
    u^{n+1}(0)=u_0 \nonumber.
\end{eqnarray}
Subsequently, we define $z_i^{n+1}$ as a~unique solution of
\begin{eqnarray}
\label{ode}
    \frac{d}{d t}{z_i}^{n+1}= \frac{1}{\beta_i} \kappa_i\big(z^{n+1}(t)\big)+
    \frac{1}{\beta_i\mathcal{H}^2(S_i(z))}\int_{S_i(0)} \big(u^{n+1}(x) +G\big)J\big(x,z^{n+1}(t)\big) d\mathcal{H}^2
    (x),    \\
     z_i^{n+1}=0 \nonumber
\end{eqnarray}
on the~maximal interval of existence $[0,T_{n+1})$, where
$T_{n+1}\leq T_n$.

We shall show that sequences $u^n$ and $z^n$ are equibounded and
equicontinuous.

We can solve the parabolic equation (\ref{par}) using methods from
previous section. Let as denote $H(t)=H(z(t))$ and $\tilde{H}(t)
=\tilde{H}(z(t),\dot{z}(t))$.

We will show that the equation (\ref{par}) has a unique solution.
Indeed, let us denote by ${\cal T}$ an operator defined as follows
\begin{eqnarray*}
    {\cal T}w (t) = U(t,0)u_0 +\int_{0}^t  U(t,\sigma) \left( \delta w(\sigma) +\tilde{H}(\sigma) \nabla w(\sigma)
    +\left(H(\sigma)
    +\tilde{H}(\sigma)\nabla - \frac{\partial}{\partial \sigma}\right)G \right) d \sigma.
\end{eqnarray*}
We will show that the~operator  ${\cal T}$ possesses exactly one
fixed point. First of all we shall show that there exist $T_n  >0$
and $\varepsilon_n >0$ such that
\begin{eqnarray*}
    B(u_0,\varepsilon_n) \subset C([0,T_n];H^s)\\
    {\cal T}: B(u_0,\varepsilon_n) \rightarrow B(u_0,\varepsilon_n).
\end{eqnarray*}
Let us notice
\begin{eqnarray*}
    \sup_{s \in [0,T_{n-1}]} |\hspace{-0.5mm}|\tilde{H}(s)|\hspace{-0.5mm}|_{L_{\infty}(\Omega^c)}
    \leq\\
     \sup_{(z,v) \in B(\varepsilon_0) \times B(\varepsilon_0')} \left|\hspace{-0.5mm}\left|\left( \frac{\partial \Lambda}
    {\partial s}\right)^{T}\left(\nabla_{(x_1,x_2,x_3)}\Lambda(x_1,x_2,x_3,s)\right)^{-1}\right|\hspace{-0.5mm}\right|_{L_{\infty}(\Omega^c)}
     +\\+\sup_{s \in
    [0,T_{1}]} \left|\hspace{-0.5mm}\left|f\left(\nabla_{(x_1,x_2,x_3)}\Lambda(x_1,x_2,x_3,s)\right)^{-1}\right|\hspace{-0.5mm}\right|_{L_{\infty}(\mathbb{R}^3 \times
\mathbb{R}_{+})} \leq M_1,
\end{eqnarray*}
where $B(\varepsilon_0)$ and $B(\varepsilon_0')$ are balls in
$C^1$ and $C^{\alpha}$ respectively, such that the solution
$z_i^0$ is~contained in $B(\varepsilon_0)$ and $\dot{z}_i^0$ in
$B(\varepsilon_0')$.

 Since $G$ has a compact support it is not
hard to~see that
\begin{eqnarray*}
    \sup_{s \in [0,T_{n-1}]} \left|\hspace{-0.5mm}\left|\left(H(\sigma)
    +\tilde{H}(\sigma)\nabla - \frac{\partial}{\partial \sigma}\right)G\right|\hspace{-0.5mm}\right|_{L_{2}(\Omega^c)} \leq M_2.
\end{eqnarray*}
Using the above inequalities and Corollary \ref{zan} we can write
\begin{eqnarray*}
    \big|\hspace{-0.5mm}\big|{\cal T}w(t)-u_0\big|\hspace{-0.5mm}\big|_{H^s} \leq c\int_0^t\big|\hspace{-0.5mm}\big|A(\sigma)U(\sigma,0)u_0\big|\hspace{-0.5mm}\big|_{\left[D(A(\sigma)),L_2\right]_{\beta}} d \sigma  +
    c \int_{0}^t \big|\hspace{-0.5mm}\big| U(t,\sigma)\delta w(\sigma)\big|\hspace{-0.5mm}\big|_{\left[D(A(t)),L_2\right]_{\beta}} d \sigma + \\
    c \int_{0}^t \big|\hspace{-0.5mm}\big| U(t,\sigma)\tilde{H}(\sigma) \nabla w(\sigma)\big|\hspace{-0.5mm}\big|_{[D(A(t)),L_2]_{\beta}} d \sigma +
    c \int_{0}^t \left|\hspace{-0.5mm}\left| U(t,\sigma)\left(H(\sigma)
    +\tilde{H}(\sigma)\nabla - \frac{\partial}{\partial \sigma}\right)G\right|\hspace{-0.5mm}\right|_{[D(A(t)),L_2]_{\beta}} d \sigma.
\end{eqnarray*}
Using the methods from \cite{Lunardi} one can show that the
following inequalities
\begin{eqnarray*}
    \big|\hspace{-0.5mm}\big|A(t)U(t,\sigma)u\big|\hspace{-0.5mm}\big|_{\left[D(A(t)),L_2\right]_{\beta}} \leq
    \frac{c}{(t-\sigma )^{1+\beta -\theta}} \big|\hspace{-0.5mm}\big|u\big|\hspace{-0.5mm}\big|_{\left[D(A(\sigma)),L_2\right]_{\theta}},\\
     \left|\hspace{-0.5mm} \left| U(t,\sigma)u \right| \hspace{-0.5mm} \right|_{[D(A(t)),L_2]_{\beta}} \leq
     \frac{c}{(t - \sigma)^{\beta}} \left|\hspace{-0.5mm}\left| u\right|\hspace{-0.5mm}\right|_{L_2}
\end{eqnarray*}
hold for $u \in D(A(\sigma))$.
Next, we obtain
\begin{eqnarray*}
    \big|\hspace{-0.5mm}\big|{\cal T}w(t)-u_0\big|\hspace{-0.5mm}\big|_{H^s} \leq c\int_0^t\frac{1}{\sigma^{1+\beta -\theta}}\big|\hspace{-0.5mm}\big|u_0\big|\hspace{-0.5mm}\big|_{[D(A(0)),L_2]_{\theta}} d \sigma  +
    c\delta \int_{0}^t \frac{1}{(t-\sigma)^{\beta}}\big|\hspace{-0.5mm}\big| w(\sigma)\big|\hspace{-0.5mm}\big|_{L_2} d \sigma + \\
    c \int_{0}^t \frac{1}{(t-\sigma)^{\beta}}\big|\hspace{-0.5mm}\big|\tilde{H}(\sigma) \nabla w(\sigma)\big|\hspace{-0.5mm}\big|_{L_2} d \sigma +
    c \int_{0}^t \frac{1}{(t-\sigma)^{\beta}}\left|\hspace{-0.5mm}\left|\left(H(\sigma)
    +\tilde{H}(\sigma)\nabla - \frac{\partial}{\partial \sigma}\right)G\right|\hspace{-0.5mm}\right|_{L_2} d \sigma \leq \\
    c T_n^{\theta-\beta}|\hspace{-0.5mm}|u_0|\hspace{-0.5mm}|_{[D(A(0)),L_2]_{\theta}} +cT_n^{1-\beta} \left( (\delta +M_1) \sup_{\sigma \in [0,T_n]}|\hspace{-0.5mm}|w(\sigma)|\hspace{-0.5mm}|_{H^s}
    +M_2\right),
\end{eqnarray*}
where $\theta > \beta$. But  $w \in B(u_0,\varepsilon_n)$, hence,
we choose $T_n$ and $\varepsilon_n$ such that
\begin{eqnarray}
\label{n1}
    \sup_{t\in [0,T_n]}\big|\hspace{-0.5mm}\big|{\cal T}w(t)-u_0\big|\hspace{-0.5mm}\big|_{H^s} \leq
    c T_n^{\theta-\beta}\big|\hspace{-0.5mm}\big|u_0\big|\hspace{-0.5mm}\big|_{[D(A(0)),L_2]_{\theta}} +\\cT_n^{1-\beta}
     \left( (\delta +M_1)\big|\hspace{-0.5mm}\big|w(\sigma)\big|\hspace{-0.5mm}\big|_{H^s}
    +M_2\right)+\varepsilon_n T_n
    ^{1-\beta}(c\delta+M_1 c) \leq \varepsilon_n.\nonumber
\end{eqnarray}
Now, we show that the~operator ${\cal T}$ is a contraction.
Indeed,
\begin{eqnarray*}
    \big|\hspace{-0.5mm}\big|{\cal T}u(t)-{\cal T}v(t)\big|\hspace{-0.5mm}\big|_{H^s} \leq c\delta \int_{0}^t \left|\hspace{-0.5mm}\left| U(t,\sigma)\big(u(\sigma) - v(\sigma)\big)\right|\hspace{-0.5mm}\right|_{[D(A(t)),L_2]_{\beta}} d\sigma + \\
    c\int_{0}^t \big|\hspace{-0.5mm}\big| U(t,\sigma)\tilde{H}(\sigma) \big(\nabla u(\sigma) - \nabla v(\sigma)\big)\big|\hspace{-0.5mm}\big|_{[D(A(t)),L_2]_{\beta}}  d\sigma \leq
    c \delta \int_{0}^t \frac{1}{(t-\sigma)^{\beta}}\big|\hspace{-0.5mm}\big|u(\sigma) - v(\sigma)\big|\hspace{-0.5mm}\big|_{L_2} d\sigma + \\ \nonumber
    c\int_{0}^t \frac{1}{(t-\sigma)^{\beta}}\big|\hspace{-0.5mm}\big|\tilde{H}(\sigma) \big(\nabla u(\sigma) - \nabla v(\sigma)\big)\big|\hspace{-0.5mm}\big|_{{L^2}} d\sigma \leq
    (c\delta + cM_1)T_n^{1-\beta} \sup_{\sigma \in[0,T_n]} \big|\hspace{-0.5mm}\big|u(\sigma)-v(\sigma)\big|\hspace{-0.5mm}\big|_{H^s}.\nonumber
\end{eqnarray*}
We find $\varepsilon_n$ and $T_n$ such that ${\cal T}$ is a
contraction and the inequality (\ref{n1}) is satisfied. Hence,
from~Banach fixed point theorem we obtain that ${\cal T}$ has a
unique fixed point. Let us notice that $\varepsilon_n =
\varepsilon_2$ and $T_n =T_2$ for $n \geq 2$.

We will show the~following
\begin{lem}
    The~sequence $u^n$ is equibounded and equicontinuous on
    $[0,T_2]$.
\end{lem}
\begin{proof}
 Equiboundedness follows from previous considerations and
 construction of the~sequence $u^n$. Let us denote by $a$ an upper
 bound of $u^n$, i.e.
\begin{eqnarray*}
    \sup_{\sigma \in [0,T_2]} |\hspace{-0.5 mm}|u^n (\sigma)|\hspace{-0.5mm}|_{H^s} \leq a.
\end{eqnarray*}

Now, we show that the sequence $u^n$ is equicontinuous. Namely,
\begin{eqnarray*}
     |\hspace{-0.5mm}|u^{n+1}(t) - u^{n+1}(\tau)|\hspace{-0.5mm}|_{H^s } \leq |\hspace{-0.5mm}| U(t,0,)u_0-  U(\tau,0)u_0|\hspace{-0.5mm}|_{H^s} +\\
      +\int_0^t \left|\hspace{-0.5mm}\left|\big(U(t,\sigma)-U(\tau,\sigma)\big)\left(\delta u^{n+1}(\sigma)+\tilde{H}(\sigma)\nabla u^{n+1} (\sigma) +\left(
      H(\sigma)+\tilde{H}(\sigma)\nabla -\frac{\partial}{\partial \sigma}\right)G \right)\right|\hspace{-0.5mm}\right|_{H^s} d \sigma +\\+
      \int_{\tau}^t \left|\hspace{-0.5mm}\left|U(t,\sigma)\left(\delta u^{n+1}(\sigma)+\tilde{H}(\sigma)\nabla u^{n+1} (\sigma) +\left(
      H(\sigma)+\tilde{H}(\sigma)\nabla -\frac{\partial}{\partial \sigma}\right)G \right)\right|\hspace{-0.5mm}\right|_{H^s} d \sigma =\\= I_1 +I_2 +I_3.
\end{eqnarray*}
It is not hard to~show the~following estimate
\begin{eqnarray*}
     I_1 \leq c |\hspace{-0.5mm}|u_0|\hspace{-0.5mm}|_{[D(A(0)),L_2]_{\theta}}\, |t-\tau|^{\theta-\beta}.
\end{eqnarray*}
Next,
\begin{eqnarray*}
     I_2 \leq \\
     c \int_0^{\tau}\hspace{-2mm} \int_{\tau }^{t }\left|\hspace{-0.5mm}\left| A( \eta)U(\eta,\sigma)\left(\delta u^{n+1}(\sigma)+\tilde{H}(\sigma)\nabla
     u^{n+1} (\sigma) +\left(
      H(\sigma)+\tilde{H}(\sigma)\nabla -\frac{\partial}{\partial \sigma}\right)G \right)\right|\hspace{-0.5mm}\right|_ {[D(A(\eta)),L_2]_{\beta}} d \eta d \sigma \leq \\
      c_1 \int_0^{\tau} \hspace{-2mm}\int_{\tau }^{t }\frac{1}{(\eta-\sigma)^{1+\beta}} \left( \delta|\hspace{-0.5mm}|u^{n+1}(\sigma)|\hspace{-0.5mm}|_{L^2}+
      |\hspace{-0.5mm}|\tilde{H}(\sigma)\nabla u^{n+1} (\sigma)|\hspace{-0.5mm}|_{L^2} +\left|\hspace{-0.5mm}\left|\left(H(\sigma)+\tilde{H}(\sigma)\nabla -\frac{\partial}{\partial
      \sigma}\right)G \right|\hspace{-0.5mm}\right|_ {L_2}\right) d \eta d \sigma \leq \\   
      \big((\delta+M_1)a+M_2\big)c_2 |t-\tau|^{1-\beta}.
\end{eqnarray*}
Subsequently, we estimate the~last term $I_3$:
\begin{eqnarray*}
     I_3 \leq c \int_{\tau }^{t }\left|\hspace{-0.5mm}\left| U(t,\sigma)\left(\delta u^{n+1}(\sigma)+\tilde{H}(\sigma)\nabla u^{n+1}
     (\sigma) +\left(
      H(\sigma)+\tilde{H}(\sigma)\nabla -\frac{\partial}{\partial \sigma}\right)G \right|\hspace{-0.5mm}\right|_ {[D(A(t)),L_2]_{\beta}}\right) d \sigma \leq \\
      c_1 \int_{\tau }^{t }\frac{1}{(t-\sigma)^{\beta}} \left( \delta|\hspace{-0.5mm}|u^{n+1}(\sigma)|\hspace{-0.5mm}|_{L^2}+
      |\hspace{-0.5mm}|\tilde{H}(\sigma)\nabla u^{n+1} (\sigma)|\hspace{-0.5mm}|_{L^2}
      +|\hspace{-0.5mm}|\left(H(\sigma)+\tilde{H}(\sigma)\nabla -\frac{\partial}{\partial \sigma}\right)G |\hspace{-0.5mm}|_ {L_2}\right) d \sigma \leq \\
      \left((\delta+M_1)a+M_2\right)c_3 |t-\tau|^{1-\beta}. 
\end{eqnarray*}
Finally,
\begin{eqnarray*}
     |\hspace{-0.5mm}|u^{n+1}(t) - u^{n+1}(\tau)|\hspace{-0.5mm}|_{H^s } \leq C\left(|t-\tau|^{\theta-\beta} +|t-\tau|^{1-\beta}\right).
\end{eqnarray*}
\end{proof}

 Now, we show the~following
\begin{lem} The~sequence $u^n$ has the~following properties
\begin{eqnarray*}
    |\hspace{-0.5mm}|u^n|\hspace{-0.5mm}|_ {H^{s} (\mathbb{R}^3 \setminus B(0,N))} \rightarrow 0
\end{eqnarray*}
 as $N \rightarrow \infty$ uniformly with respect to~$n$.
\end{lem}
\begin{proof}
Let us denote by $B^c_N = \mathbb{R}^3 \setminus B(0,N)$. First of
all we define the smooth map $\Phi_N : \Omega^c \rightarrow
[0,1]$ as follows
 \[
 \Phi_N (r) =
  \begin{cases}
      1 & \textrm{ if } r \geq N\\
      0 & \textrm{ if } r \leq \frac{N}{2},
 \end{cases}
 \]
 where $r^2 =x_1^2 +x_2^2 +x_3^2$. We require that function
 $\Phi_N$ fulfill the following condition
\begin{eqnarray*}
    |\Phi^{(k)} (r)| \leq \frac{c}{N^k}.
\end{eqnarray*}
Subsequently, we define the map $u^n_N := \Phi_N u^n$. One can
easily show that $u^n_N$ solves the problem
\begin{eqnarray*}
    \frac{\partial u^n_N}{\partial t} = H u^n_N+  \tilde{H}\nabla (\Phi_N u^n) -
    u^n \tilde{H} \nabla \Phi_N+\Phi_N
    \left(H+\tilde{H}\nabla
    -\frac{\partial}{\partial t} \right)G -\\- H \Phi_N u^n - 2(\tilde{H}-f)\nabla \Phi_N (\tilde{H} - f) \nabla u^n
\end{eqnarray*}
with initial condition $u^n_N (0) = u_0^N= \Phi_N u_0$. It is
enough to show that $|\hspace{-0.5mm}|u_N^n|\hspace{-0.5mm}|_
{H^{s} (\Omega^c)} \rightarrow 0$ as $N \rightarrow \infty$
uniformly with respect to~$n$.

Let us notice that
\begin{eqnarray*}
|\hspace{-0.5mm}|U(t,0)u^N_0|\hspace{-0.5mm}|_{H^s(\Omega^c)} \leq
|\hspace{-0.5mm}|u^N _0|\hspace{-0.5mm}|_{H^s(\Omega^c)} +c_1 T_2
^{\theta
-\beta}|\hspace{-0.5mm}|u^N_0|\hspace{-0.5mm}|_{[D(A(0)),L_2]_{\theta}(\Omega^c)}.
\end{eqnarray*}
Now, we obtain the following estimate
\begin{eqnarray*}
    \sup_ {\sigma \in [0,T_2]}|\hspace{-0.5mm}| u_N^{n}(\sigma)|\hspace{-0.5mm}|_{H^s(\Omega^c)} \leq
     c|\hspace{-0.5mm}|u_0^N|\hspace{-0.5mm}|_{H^s(\Omega^c)} +c
     T_2^{\theta
     -\beta}|\hspace{-0.5mm}|u^N_0|\hspace{-0.5mm}|_{[D(A(0)),L_2]_{\theta}(\Omega^c)}+\\
     + c T_2^ {1-\beta} \left( \sup_ {\sigma \in [0,T_2]}|\hspace{-0.5mm}| u^{n}_N(\sigma)|\hspace{-0.5mm}|_{H^s(\Omega^c)} + \sup_ {\sigma \in [0,T_2]} |\hspace{-0.5mm}| \left(H(\sigma)+\tilde{H}(\sigma)\nabla -\frac {\partial}{\partial
    \sigma}\right)G|\hspace{-0.5mm}|_{L^2(B^c_{\frac{N}{2}})}\right) +\\+ \left(\frac{1}{N^2} +
     \frac{1}{N} \right)\sup_ {\sigma \in [0,T_2]}|\hspace{-0.5mm}| u^{n}(\sigma)|\hspace{-0.5mm}|_{H^s(B^c_{\frac{N}{2}})} . 
\end{eqnarray*}
Hence,
\begin{eqnarray*}
    \sup_ {\sigma \in [0,T_2]}|\hspace{-0.5mm}| u_N^{n}(\sigma)|\hspace{-0.5mm}|_{H^s(\Omega^c)} \leq
     c |\hspace{-0.5mm}|u_0^N|\hspace{-0.5mm}|_{H^s(\Omega^c)} +c
     |\hspace{-0.5mm}|u^N_0|\hspace{-0.5mm}|_{[D(A(0)),L_2]_{\theta}(\Omega^c)}+\\
     + c \sup_ {\sigma \in [0,T_2]} |\hspace{-0.5mm}| \left(H(\sigma)+\tilde{H}(\sigma)\nabla -\frac {\partial}{\partial
    \sigma}\right)G|\hspace{-0.5mm}|_{L^2(B^c_{\frac{N}{2}})}+ \left(\frac{1}{N^2} +
     \frac{1}{N} \right)\sup_ {\sigma \in [0,T_2]}|\hspace{-0.5mm}| u^{n}(\sigma)|\hspace{-0.5mm}|_{H^s(B^c_{\frac{N}{2}})} . 
\end{eqnarray*}
From above inequality and $\sup_{\sigma \in [0,T_2]} |\hspace{-0.5
mm}|u^n (\sigma)|\hspace{-0.5mm}|_{H^s} \leq a$ the~lemma follows.
\end{proof}
Now, we examine the properties of the~sequence $z^n$. Let us
recall that $z^{n+1}$ is a~solution of~the~problem
\begin{eqnarray*}
    \frac{d}{d t}{z_i}^{n+1}=\frac{1}{\beta_i} \kappa_i(z^{n+1}(t))+\frac{1}{\beta_i\mathcal{H}^2(S_i(z))}\int_{S_i(0)} \Big(G+
    u^{n+1}(x)\Big) J(x,z^{n+1}(t)) d\mathcal{H}^2 (x), \\
     z_i^{n+1}=0.
\end{eqnarray*}
First of all we show that the above problem has a unique solution.
We define the~operator~${\cal F}$ as~follows
\begin{eqnarray*}
    {\cal F}z_i(t) =\frac{1}{\beta_i} \int_0 ^t \kappa_i (z(s)) ds + \int_0 ^t \frac {1}{\beta_i\mathcal{H}^2(S_i(z))}\int_{S_i(0)}\Big(G+
    u^{n+1}(x)\Big)
    J(x,z)(t)) d\mathcal{H}^2 (x)ds.    
\end{eqnarray*}
We will show that there exist $\varepsilon_{n+1}$ and $T_{n+1}$ such that
\begin{eqnarray*}
    B(0,\varepsilon_{n+1}) \subset C\left([0,T_{n+1});\mathbb{R}^{N+2}\right)\\
    {\cal F}: B(0,\varepsilon_{n+1}) \rightarrow B(0,\varepsilon_{n+1})
\end{eqnarray*}
and $F$ is a contraction on $B(0,\varepsilon_{n+1})$. Let us compute
\begin{eqnarray*}
    |{\cal F}z(t)-0| \leq  \frac{1}{\beta_i}\int_0 ^{T_{n+1}} \Big|\kappa_i (z(s))\Big|ds  + \frac{K}{\beta_i}\int_0 ^{T_{n+1}}
    \frac{1}{\mathcal{H}^2(S_i(z))}\int_{S_i(0)} \Big|u^{n+1}(x) +G\Big|
    d\mathcal{H}^2 (x)ds \\ \leq \frac{T_{n+1}}{\beta} \left(c_1 + \frac{K}{k} \gamma \mathcal{H}^2 (\partial \Omega^c) a~+ \frac{K}{k}
    b\right),
\end{eqnarray*}
where
\begin{eqnarray*}
    c_1 =\sup_{1 \leq i \leq N+2} \sup_{z \in B(0,\varepsilon_2)} |\kappa_i(z)|,\\
    \beta =\inf_{1 \leq i \leq N+2} \beta_i,\\
    K=\sup_{x \in \partial \Omega^c} \sup_{z\in B(0, \varepsilon_2)}|J(x,z)|,\\
    k=\inf_{1 \leq i \leq N+2}\inf_{z\in B(0, \varepsilon_2)} \mathcal{H}^2(S_i(z)),\\
    b = \sup_{1 \leq i \leq N+2} \sup_{z \in B(0,\varepsilon_2)}\int_{S_i(0)} |G|d\mathcal{H}^2 (x)
\end{eqnarray*}
and $\gamma$ is the constant from the trace theorem.

 Now, we show
that ${\cal F}$ is contraction on $B(0,\varepsilon_{n+2})$.
\begin{eqnarray*}
    \Big|{\cal F} z(t)-{\cal F}\tilde{z}(t)\Big| \leq  \frac{1}{\beta}\int_0 ^{T_{n+1}} \Big|\kappa_i (z(s))-\kappa_i (\tilde{z}(s))\Big|ds  +\\
    +\frac{1}{\beta}\int_0 ^{T_{n+1}}  \int_{S_i(0)}\left|(u^{n+1}(x) +G)\left( \frac{J(z(s))}{\mathcal{H}^2(S_i(z))}
    -\frac{J(\tilde{z}(s))}{\mathcal{H}^2(S_i(\tilde{z}))}\right)\right| \leq \\
   \frac{1}{\beta} \left(c +k c_2 \gamma a+ c_2 b\right)T_{n+1}\sup_{s \in [0,T_{n+1}]}|z(s)-\tilde{z}(s)|.
\end{eqnarray*}
Hence, form contraction principle the solvability of our problem
follows. Let us notice that $\varepsilon_n = \varepsilon_2$ and
$T_n =T_2$ for $n \geq 2$.

We have the~following
\begin{lem} The~sequences $z^n$ and $\dot{z}^n$ are equibounded and equicontinuous.
\end{lem}
\begin{proof}
From above consideration follows that $z^{n}$ is equibounded. We show that it is equicontinuous. Namely,
\begin{eqnarray*}
    \Big|z^{n+1}(t)-z^{n+1}(\sigma)\Big| \leq \frac{1}{\beta} \int_{\sigma} ^t \Big| \kappa_i (z(s))\Big| ds +\\
    + \frac{1}{\beta}\int_{\sigma} ^t \frac{1}{\mathcal{H}^2(S_i(z^{n+1}))}\int_{S_i(0)} \Big|u^{n+1}(x)
    J(x,z)(t)\Big| d\mathcal{H}^2 (x)ds + \\
    +\frac{1}{\beta}\int_{\sigma} ^t \frac{1}{\mathcal{H}^2(S_i(z^{n+1}))}\int_{S_i(0)} \Big|G
    J(x,z)(t)\Big| d\mathcal{H}^2 (x)ds  \leq \frac{1}{\beta}\left(c_1 +\frac{K}{k}\gamma a+\frac{K}{k}b\right) |t-\sigma|.
\end{eqnarray*}
Subsequently, we show that $\dot{z}_{n+1}$ is equibounded,
\begin{eqnarray*}
    \left|\dot{z}^{n+1}\right| \leq \frac{1}{\beta}\left(c_1+\frac{K}{k} \gamma a+\frac{K}{k}b \right).
\end{eqnarray*}
Next, equicontinuouity follows from the estimate
\begin{eqnarray*}
    \left|\dot{z}^{n+1}(t) - \dot{z}^{n+1}(\sigma)\right| \le \frac{1}{\beta}\left|\kappa_i (z^{n+1}(t))-\kappa_i (z^{n+1}(\sigma))\right|+\\
    +\frac{1}{\beta}\int_{S_i(0)}\left|(u^{n+1}(t) +G(t)) \frac{J(z^{n+1}(t))}{\mathcal{H}^2(S_i(z^{n+1}))}
    -\left(u^{n+1}(\sigma)+ G(\sigma)\right)\frac{J(z^{n+1}(\sigma))}{\mathcal{H}^2(S_i(z^{n+1}))}\right| d\mathcal{H}^2 \leq c|t-\sigma|^{\alpha},
\end{eqnarray*}
where we applied Lipschitz continuity of $z$ and H\"older continuity of $u^{n+1}$.
\end{proof}
From the above three lemmas, Lemma \ref{zwartosc} and Ascoli Theorem the~proof follows.
\end{proof}
\section{Summary}
In this paper we have considered the system modelling evolution of
crystals 3-D from supersaturated vapor. We have shown local in
time existence of solutions. Let us mention that we could not
expect global in time existence of solutions (topological
catastrophe could occur). Let us stress that uniqueness of
solutions is still an open problem. In the forthcoming article
(see \cite{Gorka}) we will show local in time existence of
solutions to the problem with Stefan conditions. In this problem
we shall apply Galerkin method and we shall obtain less regular
solutions.
\\
\\
\textbf{Acknowledgement.}\\
\\
Author would like to thank Professor Piotr Rybka for his invaluable
comments, suggestions and reading preliminary version of this
paper. He would also like to thank his wife Małgosia for drawing figures contained
in this paper. The author was in part supported by KBN grant 1 P03A 37 28.
\section{Appendix}
This section contains an auxiliary and technical result.
\begin{lem}
\label{rozszerzenie} Let us assume that $\Omega$ is an admissible
shape, i.e. $\Omega = W \times [-L,L]$. Then there exist
continuous maps $h_i$ ($i=1,2,...,N+2$) with compact support such
that $\frac{\partial  h_i}{\partial x_l}$ and $\frac{\partial ^2
h_i}{\partial x_k \partial x_l}$ exist a.e. and $\frac{\partial
h_i}{\partial x_l} , \frac{\partial ^2 h_i}{\partial x_k \partial
x_l} \in L^2(\Omega^c)$  ($i=1,2,...,N+2$ and $k,l = 1,2,3$).
Moreover the~following is~true
\begin{eqnarray*}
    \frac{ \partial h_i} {\partial \nu_j } \Big|_ { { S_j (0)}} = \delta_{i,j}.
\end{eqnarray*}
\end{lem}
\begin{proof}
It is enough to~construct maps $h_i$ for $i=2,T,B$.

Let us introduce $N+2$ cut-off maps  $\eta_i \in C^{\infty}_0
(\mathbb{R}^3)$, $0\leq \eta_i \leq 1$. The~following positions of
lateral facets are possible (see Figure 7)
\begin{center}

\includegraphics[scale=.45]{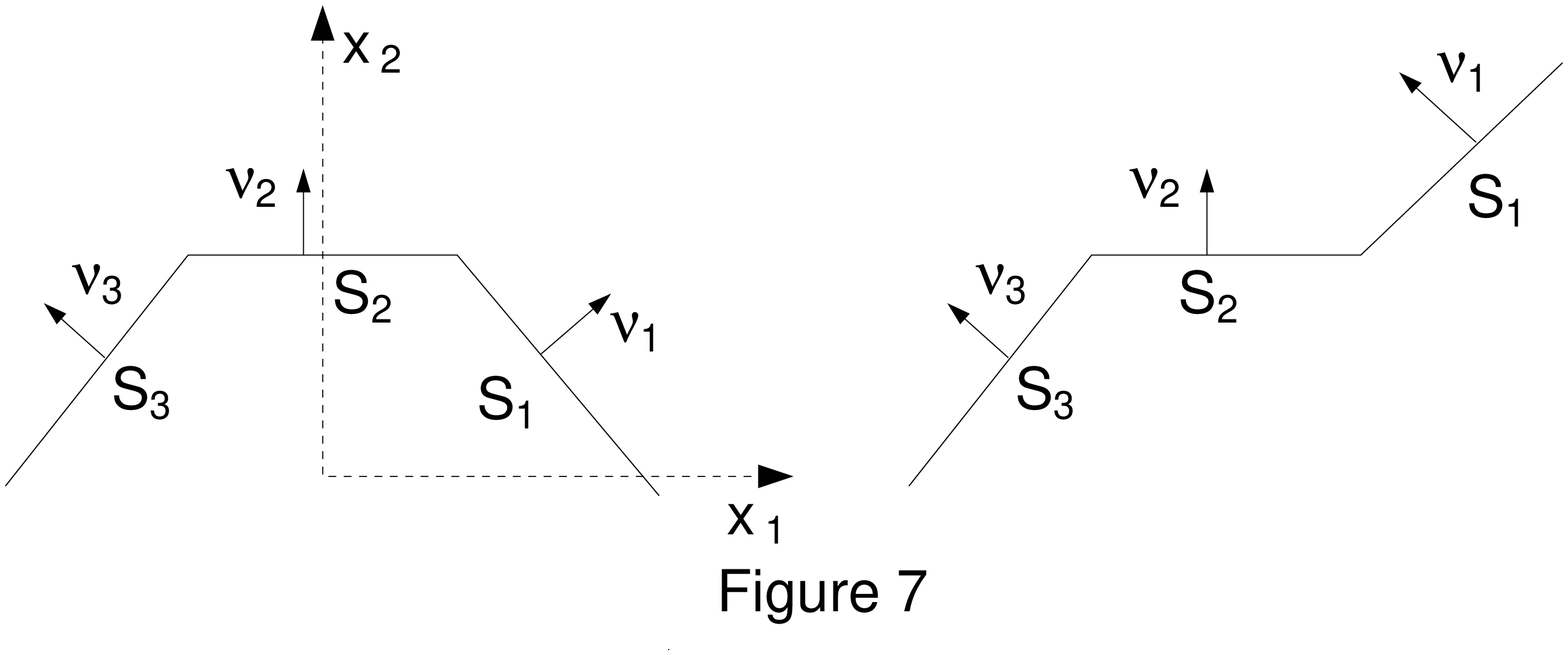}

\end{center}

Let us denote $\delta =10 diam(\Omega)$. For $i=1,2,...,N$ we
define cut-off function as follows
\begin{displaymath}
    \eta_i (x_1,x_2, x_3)=\left\{ \begin{array}{ll} 0 & \textrm{ if $(x_1,x_2) \notin (L_i)_{4 \varepsilon}$
     or $|(x_1,x_2, x_3)| \geq \delta$}\\
     1 & \textrm{ if $(x_1,x_2) \in
     (L_i)_{\varepsilon} $ and $|x_3 | \leq \frac{\delta}{2}$,}
      \end{array} \right.
\end{displaymath}
where $\varepsilon$ is sufficiently small. Subsequently
\[
\eta_T (x_1,x_2, x_3)=
\begin{cases}

 0 & \textrm{if $x_3 \geq 0$     or $|(x_1,x_2, x_3)| \geq \delta$}\\
     1 & \textrm{ if $x_3 \in [L-\varepsilon, L + \varepsilon]$ and $|(x_1 , x_2)| \leq \frac{\delta}{2}$}
\end{cases}
\]
and $\eta_B (x_1,x_2, x_3) = \eta_T (x_1,x_2, -x_3)$. Next we denote
\begin{eqnarray*}
     h_T (x_1,x_2, x_3)= \eta_T (x_1,x_2, x_3) x_3 ,\\
     h_B (x_1,x_2, x_3)= \eta_B (x_1,x_2, x_3) x_3 .
\end {eqnarray*}
and
\[
h_2 (x_1,x_2, x_3)=
\begin{cases}
     x_2 \eta_2 (x_1,x_2, x_3)& \textrm{ if }(x_1,x_2) \in A_2\\
     x_2 \eta_2 (x_1,x_2, x_3) \phi_2 (x_1) & \textrm{ if }(x_1,x_2) \in B_2\\
     x_2 \eta_2 (x_1,x_2, x_3) \phi_2 '(x_1) & \textrm{ if }(x_1,x_2) \in C_2,
\end{cases}
\]
where the sets $A_2 , B_2$ and $C_2$ are explained in Figure 8.
\begin{center}

\includegraphics[scale=.4]{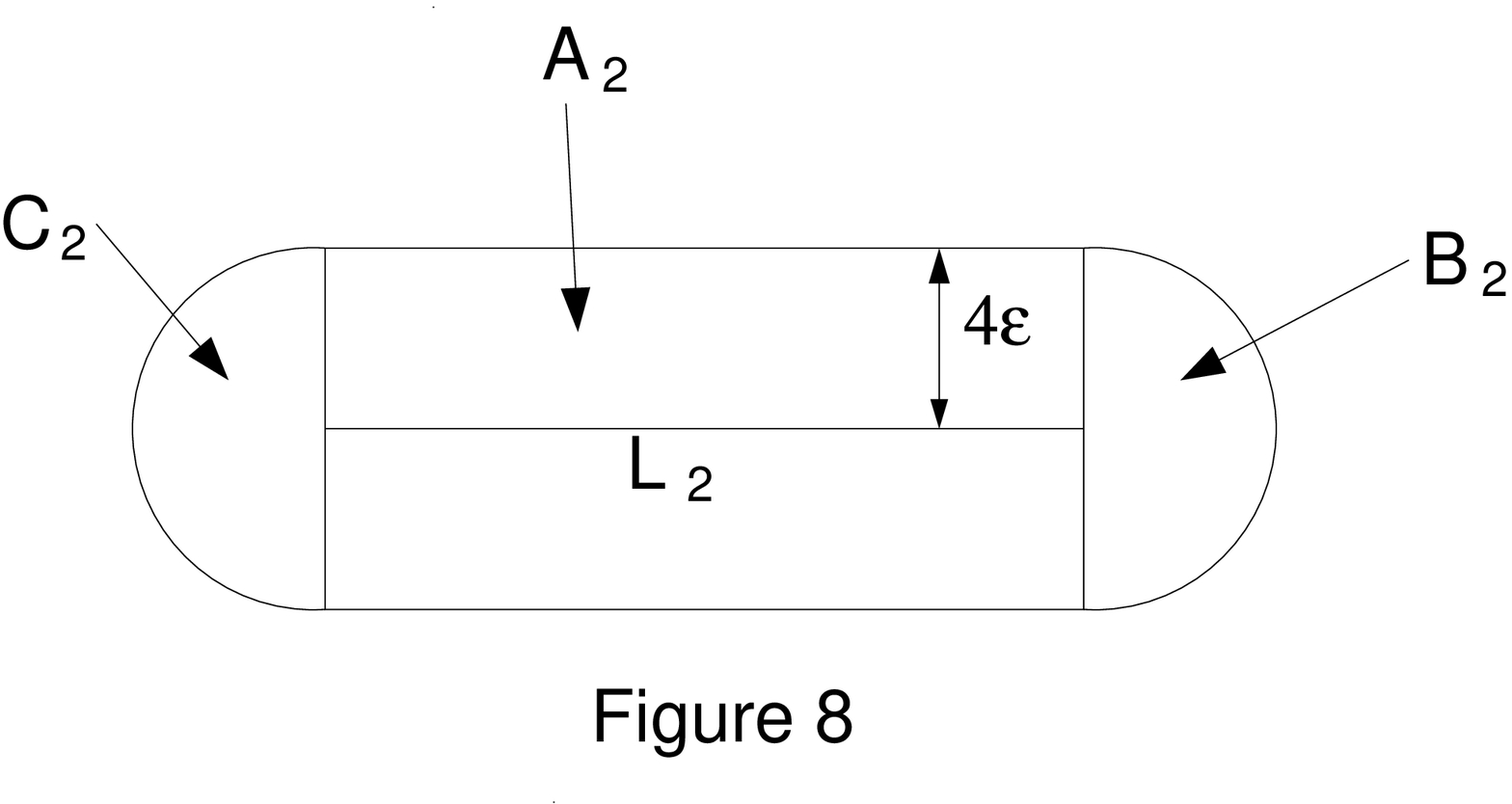}

\end{center}

where $\phi_2 (x_1)$ and $\phi_2 '(x_1)$ are chosen in such a way
that the required properties are fulfilled.

\end{proof}

\end{document}